\documentclass[twocolumn,prd,aps,superscriptaddress,preprintnumbers,tightenlines,showpacs,nofootinbib,eqsecnum,amsfonts,amsmath]{revtex4-1}
\usepackage[normalem]{ulem}
\usepackage{placeins}
\usepackage{color}
\usepackage{calc}
\usepackage{amsmath,amssymb,graphicx}
\usepackage{tensor}
\usepackage{bm}
\usepackage{times}
\usepackage[varg]{txfonts}
\usepackage[colorlinks, pdfborder={0 0 0}]{hyperref}
\usepackage{float}
\usepackage{dcolumn}
\usepackage[nolist,nohyperlinks]{acronym}
\usepackage{xspace}
\usepackage[abs]{overpic}
\usepackage{pict2e}
\usepackage{enumitem}
\usepackage[usenames,dvipsnames]{xcolor}
\usepackage[utf8]{inputenc}
\usepackage{acronym}
\usepackage{gensymb}
\usepackage{cleveref}
\usepackage[normalem]{ulem}
\usepackage{longtable}
\usepackage{verbatim}
\usepackage{multirow}

\usepackage{subfigure}
\usepackage{placeins}
\usepackage{hyperref}
\usepackage{aas_macros}

\definecolor{rb4}{HTML}{27408B}

\begin{document}

\title{Discriminating between different scenarios for the formation and evolution of massive black holes with LISA }

\author{Alexandre Toubiana}
\affiliation{Université de Paris, CNRS, Astroparticule et Cosmologie,  F-75006 Paris, France}
\affiliation{Institut d'Astrophysique de Paris, CNRS \& Sorbonne
 Universit\'es, UMR 7095, 98 bis bd Arago, 75014 Paris, France} 

\author{Kaze W.K. Wong}
\affiliation{Department of Physics and Astronomy, Johns Hopkins University, 3400 N. Charles Street, Baltimore, Maryland 21218, USA}

\author{Stanislav Babak}
\affiliation{Université de Paris, CNRS, Astroparticule et Cosmologie,  F-75006 Paris, France}
\affiliation{Moscow Institute of Physics and Technology, Dolgoprudny, Moscow region, Russia}

\author{Enrico Barausse}
\affiliation{SISSA, Via Bonomea 265, 34136 Trieste, Italy \& INFN,
Sezione di Trieste}
\affiliation{IFPU - Institute for Fundamental Physics of the Universe,
Via Beirut 2, 34014 Trieste, Italy}

\author{Emanuele Berti}
\affiliation{Department of Physics and Astronomy, Johns Hopkins University, 3400 N. Charles Street, Baltimore, Maryland 21218, USA}

\author{Jonathan R. Gair}
\affiliation{Max Planck Institute for Gravitational Physics (Albert Einstein Institute), Am M¨uhlenberg 1, Potsdam-Golm, 14476, Germany}
\affiliation{School of Mathematics, University of Edinburgh, James Clerk Maxwell Building, Peter Guthrie Tait Road, Edinburgh EH9 3FD, United Kingdom}

\author{Sylvain Marsat}
\affiliation{Université de Paris, CNRS, Astroparticule et Cosmologie,  F-75006 Paris, France}

\author{Stephen R. Taylor}
\affiliation{Department of Physics \& Astronomy, Vanderbilt University,
2301 Vanderbilt Place, Nashville, Tennessee 37235, USA}

\begin{abstract}
  Electromagnetic observations have provided strong evidence for the existence of massive black holes in the center of galaxies, but their origin is still poorly known. Different scenarios for the formation and evolution of massive black holes lead to different predictions for their properties and merger rates. LISA observations of coalescing massive black hole binaries could be used to reverse engineer the problem and shed light on these mechanisms. In this paper, we introduce a pipeline based on hierarchical Bayesian inference to infer the mixing fraction between different theoretical models by comparing them to LISA observations of massive black hole mergers. By testing this pipeline against simulated LISA data, we show that it allows us to accurately infer the properties of the massive black hole population as long as our theoretical models provide a reliable description of the Universe. We also show that measurement errors, including both instrumental noise and weak lensing errors, have little impact on the inference. 
\end{abstract}

\maketitle

\section{Introduction}

The detection of gravitational waves in the 10-1000 Hz band over the last six years by the LIGO/Virgo collaboration~\cite{Abbott:2016blz,LIGOScientific:2018mvr,LIGOScientific:2020ibl} has allowed us to infer for the first time the population of stellar-mass black hole (BH) binaries in the Universe~\cite{LIGOScientific:2018jsj,Abbott:2020gyp}, shedding some light on their possible formation channels (see e.g.~\cite{Zevin:2017evb,Talbot:2017yur,Belczynski:2017gds,Talbot:2018cva,Roulet:2018jbe,Bouffanais:2019nrw,Baibhav:2020xdf,Roulet:2020wyq,Hall:2020daa,Wong:2020ise,Kimball:2020qyd,Zevin:2020gbd,Bouffanais:2021wcr,DeLuca:2021wjr,Gayathri:2021xwb,Franciolini:2021tla,Callister:2021fpo}). 
% ~\cite{Fishbach:2017dwv,Farr:2017uvj,Zevin:2017evb,Talbot:2017yur,Belczynski:2017gds,Rodriguez:2016vmx,Talbot:2018cva,Roulet:2018jbe,Wong:2020ise,Zevin:2020gbd,Bouffanais:2021wcr,DeLuca:2021wjr,Franciolini:2021tla}.
Scheduled for 2034, the Laser Interferometer Space Antenna (LISA)~\cite{Audley:2017drz} will be sensitive to gravitational waves in the mHz band, and will reveal a virtually unexplored population of compact binaries. Some of the anticipated sources include Galactic binaries, which will be so numerous that they will form a stochastic foreground dominating over instrumental noise but should also include $\approx 10^4$ individually resolvable binaries~\cite{Nelemans:2001hp,Korol:2017qcx}, and massive black hole binaries (MBHBs) with total mass in the range $10^4$-$10^9 \ M_{\odot}$~\cite{Sesana:2007sh,Sesana:2010wy,Klein:2015hvg,Bonetti:2018tpf,Barausse:2020mdt,Barausse:2020gbp}.

Electromagnetic observations indicate that massive BHs (MBHs) are present in the centers of most galaxies in the local  universe~\cite{1984ApJ...278...11G,Kormendy:1995er,2011Natur.470...66R,Reines:2013pia, Baldassare:2019yua}, including our own Galaxy~\cite{1999ApJ...524..816R,Schodel:2002vg,Reid_2003,Gillessen:2008qv} and M87~\cite{Akiyama:2019cqa}, and that their properties are correlated with those of their host galaxies, suggesting a  synergistic growth~\cite{Kormendy:1995er,Ferrarese:2000se,McConnell:2012hz,Schramm:2012jt,Kormendy:2013dxa}.
Unfortunately, these observations are sensitive only to active MBHs up to $z \sim 7$ (cf. e.g.~\cite{Wang_2021}), or local ones for which we can observe the gas/stellar dynamics. Gravitational waves will allow us to probe much more distant MBHs: LISA will be capable of detecting MBHBs up to $z\sim 20$, provided that they exist at such high redshift~\cite{Audley:2017drz}.
In this paper, we address the question of how these observations can help constrain scenarios for the formation and subsequent evolution of MBHs. 
 
The population of MBHBs that LISA will observe is the result of a complex evolutionary path, whose details are still largely unknown. Two open issues, of particular importance for LISA, can be highlighted. First, which astrophysical mechanisms provided the seeds that grew into MBHs? Several scenarios have been proposed, suggesting seed masses ranging from $10^2$ to $10^5$ $M_{\odot}$, forming at $z\sim 15-20$ (see e.g.~\cite{Latif:2016qau} for a review). Once these intermediate mass BHs form, they are thought to grow via gas accretion and successive mergers. Following the merger of two galaxies hosting a BH at their center, dynamical friction drives the BHs to the center of the newly formed galaxy, where they may form a bound binary system~\cite{Begelman:1980vb} (see however Ref.~\cite{Tremmel2018} for the possibility that a significant fraction of galaxy mergers may never produce a bound MBHB).  If this happens (at $\sim$ pc separation for systems of $\sim 10^8 M_\odot$), dynamical friction becomes inefficient and other processes take over to control the binary's evolution, including three body interactions with stars (stellar hardening)\cite{Quinlan:1996vp,Sesana:2015haa}, gas-driven migration~\cite{Macfadyen:2006jx,Cuadra:2008xn,Lodato_2009,Roedig_2011,Nixon:2010by,Duffell:2019uuk,Munoz:2018tnj} or interactions with other MBHs~\cite{Bonetti:2017lnj,Bonetti:2018tpf,Barausse:2020mdt}. The efficiency of these processes is uncertain, but they are crucial because it is not until $\sim 10^{-2}$ pc separations that gravitational wave emission is sufficient to make the binary coalesce within a Hubble time.  Whether MBHBs can transition efficiently from pc to sub-pc separation is therefore still uncertain, which is usually referred to in the literature as the ``last parsec problem''~\cite{Begelman:1980vb}. The physics of BH seeding at high redshift and the last parsec problem significantly affect the properties of the population of events that LISA will observe, such as the component masses and spins, the redshift, and the rates themselves. Thus, by accumulating observations with LISA, one can in principle reverse engineer the problem, and shed light on these mechanisms.
 
We focus here on the ability of LISA to distinguish between different seeding scenarios. We improve upon Refs.~\cite{Gair:2010bx,Sesana:2010wy} in a number of ways. 
We use a more refined treatment of selection effects; we use updated astrophysical models, with improved treatment 
of the baryonic physics, of the formation of MBH pairs, of the hardening of MBHBs
and of the effect of SN winds and accretion on MBH evolution; and we use more realistic assumptions about the LISA data, including an up to date model of the LISA instrument, and more realistic models for the gravitational waveforms generated by merging MBHs.
We use the predictions of the semianalytic model of Ref.~\cite{Barausse:2012fy} (with updates described in Refs.~\cite{Sesana:2014bea,Antonini:2015cqa,Antonini:2015sza,Bonetti:2018tpf,Barausse:2020mdt}) for the evolution of galaxies and MBHs to simulate LISA data. This model has light seed (LS) and heavy seed (HS) variants, differing in the prescription for the initial masses of BHs. We consider the possibility that the population of MBHs is described by a mixture between the LS and HS scenarios. We treat the mixing fraction between models as a {\it hyperparameter} controlling the population, and estimate it from simulated datasets using a hierarchical Bayesian framework. We test the robustness of our analysis by using the predictions of different semianalytic simulations to generate data, and assess the impact of measurement errors (due to detector noise and weak lensing) on our inference of the MBHB population.
 
This paper is organized as follows. In Sec.~\ref{da_pe} we explain how LISA data is simulated and how we perform parameter estimation. Sec.~\ref{mbh_cats} describes the astrophysical models used for the population of MBHs and our mixing procedure. In Secs.~\ref{sec_hba} and \ref{kde} we review the main aspects of the hierarchical Bayesian analysis and how to combine it with results from numerical simulations. We present our main results in Sec.~\ref{results} and our conclusions in Sec.~\ref{ccls}.

\section{Data simulation and parameter estimation}\label{da_pe}

LISA will observe the last stages of the coalescence of MBHBs, where higher harmonics can be comparable in amplitude to the $(2,\pm2)$ harmonics~\cite{Arun:2007hu,Trias:2007fp,Porter:2008kn,McWilliams:2009bg,Marsat:2020rtl}. Therefore, we use the phenomenological approximant PhenomHM~\cite{London:2017bcn} to generate the signal and perform parameter estimation. In this work we consider, for simplicity, quasicircular binaries with component spins aligned or antialigned with the orbital angular momentum (we comment on this in Sec.~\ref{mbh_cats}). We compute the full LISA response and parametrize MBHBs as described in~\cite{Marsat:2018oam, Marsat:2020rtl}. 
Denoting by $m_1$ and $\chi_1$ the mass and spin of the heaviest BH in a binary and by $m_2$ and $\chi_2$ those of its companion, we define the chirp mass as $\mathcal{M}_c=(m_1 m_2)^{3/5}/(m_1+m_2)^{1/5} $, the mass ratio as $q=m_1/m_2 \geq 1$ and the symmetric mass ratio as $\eta=q/(1+q)^2$. We also introduce the effective spin $\chi_+$ and the corresponding antisymmetric combination $\chi_-$, defined as $\chi_{+,-}=(m_1\chi_1 \pm m_2\chi_2)/(m_1+m_2)$. We adopt the cosmological parameters reported by the Planck mission (2018)~\cite{Planck:2018vyg} to compute the luminosity distance $D_L$ from the cosmological redshift $z$. Recall that source-frame (subscript $s$) and detector-frame (subscript $d$) masses are related via $m_{d}=(1+z)m_s$. We use the SciRDv1 noise curve~\cite{scirdv1}, including the confusion noise due to Galactic binaries~\cite{Mangiagli:2020rwz}, and assume a low-frequency cutoff of $10^{-5} \ {\rm Hz}$ in the LISA noise power spectral density. We assume a mission duration of four to ten years and an ideal $100 \%$ duty cycle. 

For our purposes we will not need state-of-the-art MBHB parameter estimation, but just realistic error estimates for the intrinsic parameters of the source and for the luminosity distance. Therefore, we work in the zero-noise approximation \cite{Rodriguez:2013oaa} and simply compute the Fisher information matrix~\cite{Vallisneri:2007ev} to obtain the errors on source parameters, and more specifically we use the extended Fisher formalism of Ref.~\cite{Toubiana:2020cqv}. A more complete parameter estimation study is in preparation. As shown in Fig.~\ref{fig:distrib_fid}, astrophysical models predict some events with large mass ratios and/or large spins, far outside the range of validity of current waveform models. Again, for simplicity, we will use PhenomHM for our calculations.

The chirp mass is the best measured parameter, and because we can observe the late inspiral and the merger-ringdown with high signal-to-noise ratio (SNR) up to thousands, we can measure the mass ratio and the spin of the primary quite accurately. For the heaviest systems, we can also measure the spin of the secondary. As for distance measurements, the error due to weak lensing dominates over the statistical error at high redshifts. We use the (pessimistic) model of~\cite{Hirata:2010ba}, which estimates that the error due to lensing goes as
\begin{equation}
\frac{\sigma_{D_L,{\rm lensing}}}{D_L}=0.066 \left [ \frac{1-(1+z)^{-0.25}}{0.25} \right ]^{1.8}.\label{lensing}
\end{equation}
We include this error by convolving the measured LISA posterior distribution with a Gaussian of width $\sigma_{D_L,{\rm lensing}}$. The error due to weak lensing propagates into the determination of source-frame masses. 

\section{Massive black holes catalogues}\label{mbh_cats}

\subsection{Semianalytic models}

To describe the expected population of MBHBs detectable by LISA, we utilize the semianalytic galaxy formation model of Ref.~\cite{Barausse:2012fy}, with updates described in Refs.~\cite{Sesana:2014bea,Antonini:2015cqa,Antonini:2015sza,Bonetti:2018tpf,Barausse:2020mdt}. Our model relies on dark matter halo merger trees produced with an extended Press-Schechter formalism~\cite{Press:1973iz}, modified to reproduce the results of N-body simulations~\cite{Parkinson2008}. Baryonic structures contained in the halos are evolved along the branches and through the nodes of these merger trees. These structures include: a diffuse intergalactic medium with primordial metallicity, which accretes onto the halos either by getting shock-heated to the halo virial temperature (in large low-redshift systems) or along cold flows (at high redshift and/or small systems)~\cite{Dekel:2004un,Cattaneo:2006rp,Schaal:2016szc}; a cold interstellar medium where star formation takes place, and which we assume to be in the form of disks and/or bulges; stellar disks and bulges; and nuclear compact configurations, i.e. nuclear star clusters and MBHs. The latter, which are obviously of crucial importance for this work, are assumed to grow from high-redshift seeds by accretion -- thus shining as quasars and active galactic nuclei (AGNs) -- and coalescences. The model also accounts for AGN feedback (i.e., the effect of AGN jets, disk winds and radiation) and supernova feedback (i.e., supernova explosions). Both processes can affect the evolution of baryonic structures, quenching star formation (mainly in large and small systems, respectively), ejecting/heating up nuclear gas, and also suppressing accretion onto MBHs.
In order to minimize the uncertainties, the model is calibrated to a number of observations at both galactic and nuclear scales~\cite{Barausse:2012fy,Sesana:2014bea,Antonini:2015sza,Antonini:2015cqa,Barausse:2017uyr,Guepin:2017abw,Barausse:2020mdt}. Nevertheless, as already mentioned, the predictions for LISA are crucially dependent on the assumptions made about two poorly understood processes: the formation of the high-redshift seeds and the ``delays'' with which MBHs come together and eventually coalesce after a galaxy merger.
  
As our fiducial astrophysical scenario, we adopt \emph{Model-delayed} of~\cite{Bonetti:2018tpf}, of which we consider two variants, with either LSs or HSs. In the LS model, MBHs grow from the remnants of Pop III stars at $z\gtrsim 15$~\cite{Madau:2001sc}. We seed large halos collapsing from the 3.5$\sigma$ peaks of the primordial
density field, and to describe the Pop III stellar mass function
we use a log-normal distribution centered at $300 M_\odot$ and with
rms of 0.2 dex (with an exclusion region between 140 and 260 $M_\odot$
to account for pair instability supernova explosions). The mass of the seed MBH
is then assumed to be $\sim2/3$ of the initial Pop III star mass, to account for the mass loss during the supernova explosion.
In the 
 HS model, MBHs form instead  with masses already $\sim 10^5$ $M_{\odot}$. In more detail, we use the model of Ref.~\cite{Volonteri:2007ax}, in which seeds form from the collapse of proto-galactic disks as a result of bar instabilities, at $z\gtrsim 15$ and
 in halos with spin parameter and virial temperature below critical threshold values. The latter are
 given by Eq.~(4) -- with $Q_c=2.5$ -- and Eq.~(5) of Ref.~\cite{Volonteri:2007ax}, and we use Eq.~(3) of the same
 work to set the seed mass. As for the delays between galaxy/halo and BH mergers, Ref.~\cite{Bonetti:2018tpf} accounts for the dynamical friction between the dark matter halos (including the effect of tidal disruption and evaporation); for the timescales associated (on much smaller $\sim$ pc scales) to stellar hardening\footnote{ As suggested by N-body simulations~\cite{Sesana:2015haa}, the stellar hardening timescales are computed 
 from the density at the
  mass influence radius of the binary, i.e. the radius at which the enclosed stellar mass is twice the binary mass.}, gas-induced migration and interactions with additional MBHs (brought in by later galaxy mergers); and finally for the gravitational-wave driven evolution timescale at sub-pc separations. The timescale associated to the binary's evolution at $\sim$kpc separations is instead neglected in Refs.~\cite{Barausse:2012fy,Bonetti:2018tpf}, on the premise that it should be negligible when compared to the other timescales involved. Recently, however, large scale cosmological simulations have challenged this notion~\cite{Tremmel2018}, i.e. they 
  have found that evolution timescales on those large separations can be significant. This prompted Ref.~\cite{Barausse:2020mdt} to include an additional timescale in the semianalytic model of Refs.~\cite{Barausse:2012fy,Bonetti:2018tpf} to account for the binary's evolution at  $\sim$kpc separations. Moreover, Ref.~\cite{Barausse:2020mdt} also modified the supernova feedback model of Refs.~\cite{Barausse:2012fy,Bonetti:2018tpf} to account for the possibility that supernova winds may quench not only star formation, but also accretion onto MBHs in low-mass, high-redshift galaxies~\cite{Habouzit2017}. We implement this effect by assuming that 
  the growth of the gas reservoir off which the MBH accretes
  is curtailed in systems with escape velocity (from the bulge) lower than 270 km/s~\cite{Habouzit2017}.
    We refer to the model including these additional ingredients (delays on scales of hundreds of pc and SN feedback on BH accretion) as \emph{SN-delays}, adopting the same designation as in Refs.~\cite{Barausse:2020mdt,Barausse:2020gbp}.

We use the semianalytic model to produce simulated populations of MBHBs, including information on their masses, spins and redshift. It is worth noting that the eccentricity of a binary and the degree of alignment of the component spins depend on the mechanism that triggers the merger. For instance, triple/quadruple interactions between MBHs can lead to large eccentricities as a result of Kozai-Lidov resonances~\cite{Kozai:1962zz,1962P&SS....9..719L} and/or chaotic interactions~\cite{Bonetti2016,Bonetti2018a}. Binaries merging in a gas-rich environment tend to have aligned spins, because of the Bardeen-Petterson effect~\cite{Bardeen:1975zz,Bogdanovic:2007hp}, i.e. the gravito-magnetic torques exerted by the circumbinary disk. We also stress that the evolution of the spin under accretion is described in our model by neither coherent nor chaotic accretion, but by the {\it hybrid} model of Ref.~\cite{Sesana:2014bea}. The latter incorporates Bardeen-Petterson torques, is intermediate between chaotic and coherent accretion, and reproduces the sample of spin measurements from iron K$\alpha$ lines.

These effects are included in our semianalytic model (cf. in particular Refs.~\cite{Barausse:2012fy,Sesana:2014bea,Bonetti:2018tpf}), with the final remnant mass and spin produced by the MBH merger computed
via fitting formulas reproducing the results of numerical-relativity simulations~\cite{Barausse:2012qz,Hofmann:2016yih}. However, the information on spin alignment and eccentricity is not fully exploited in the analysis performed for this paper. Indeed, because PhenomHM covers only quasicircular binaries with component spins aligned or antialigned with the orbital angular momentum, we simply take the projection of spins along the orbital angular momentum and neglect the eccentricity. Nevertheless, the information on the spin alignment is partially contained in the effective spin of the binary. To complete the set of parameters $\theta$ needed to describe LISA events, we draw the sky location uniformly on the sphere, the phase at coalescence and the polarization uniformly in $[0,2\pi]$, and the inclination angle $\cos \iota$ uniformly in $[-1,1]$. We assume a time to coalescence of at most one year, and we do not consider the part of the signal below $10^{-5} \ {\rm Hz}$.

\subsection{Population properties}

\begin{figure*}[hbtp]
 \centering
\subfigure[ \ Without SNR threshold.]{
    \centering \includegraphics[scale=0.19]{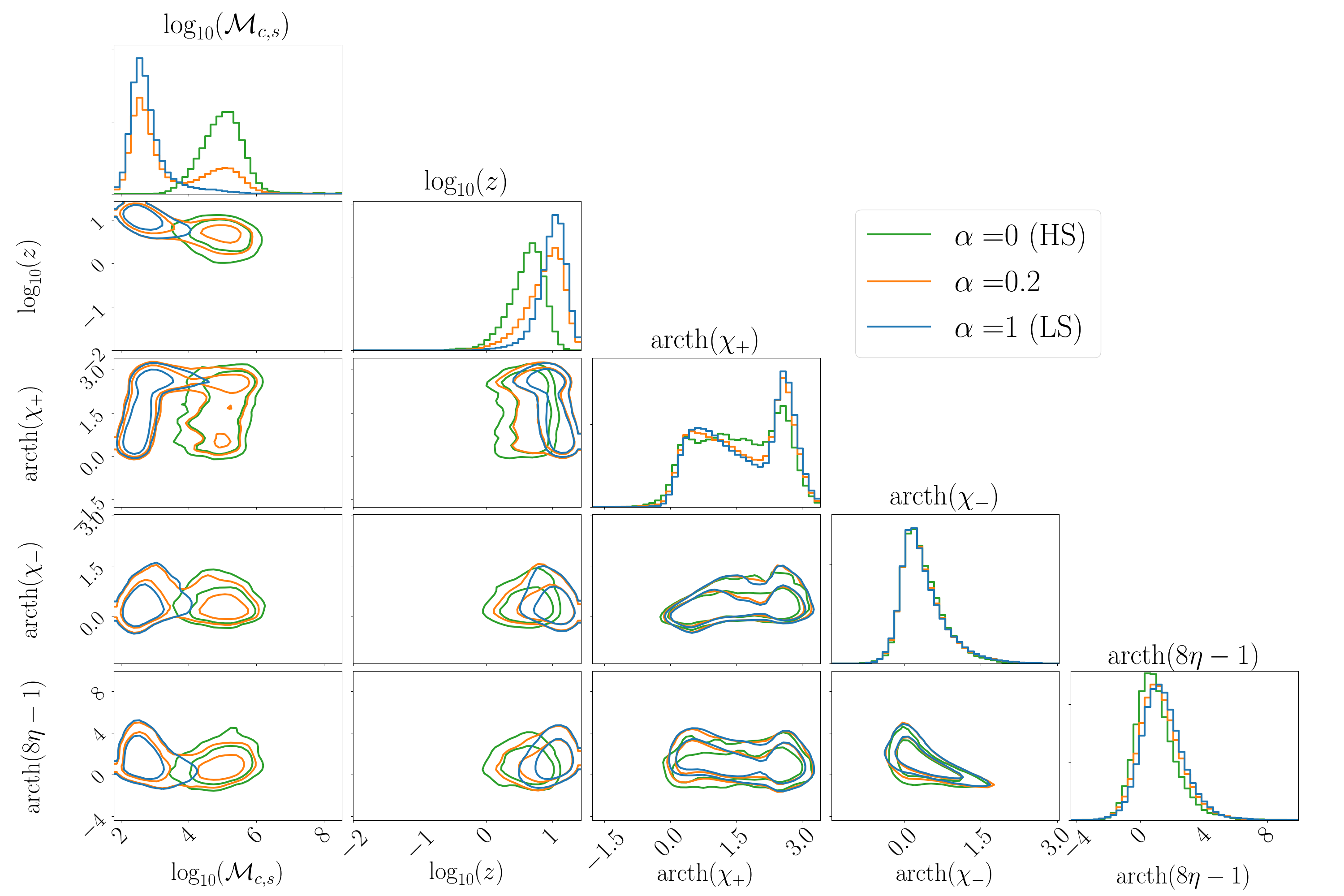}
    }
\centering
\subfigure[ \ With an SNR threshold of 10.]{
    \centering \includegraphics[scale=0.19]{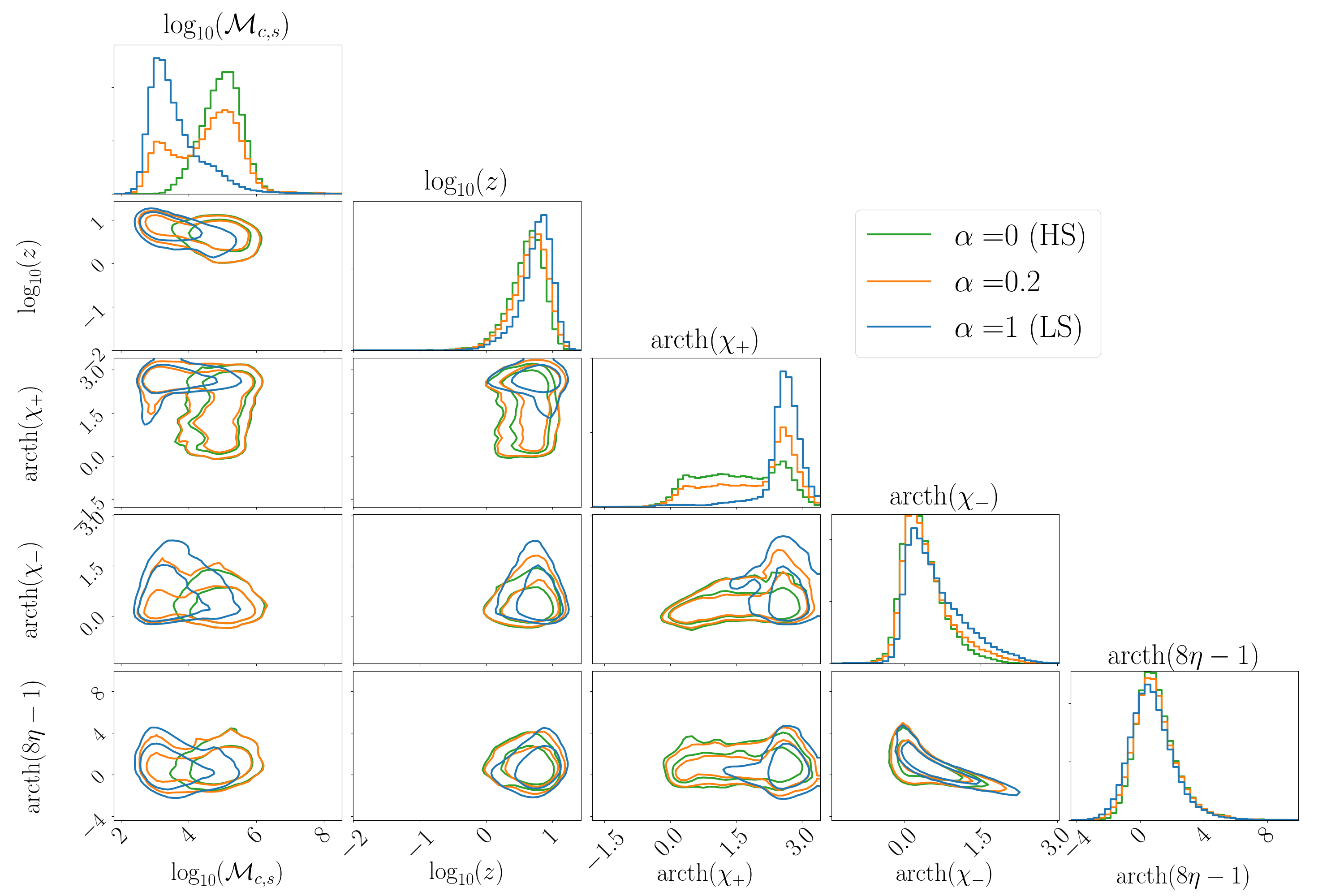}
   }
   \centering
   \caption{Normalized population distribution for different values of the mixing fraction between the fiducial LS and HS models. We show the 68\% and 90\% confidence intervals. The (source-frame) chirp mass distribution is the most sensitive to $\alpha$. The redshift distributions of detectable events look much more similar, unlike the effective spin distributions, as discussed in the main text.}\label{fig:distrib_fid}   
 \end{figure*}
 
   \begin{figure*}[hbtp]
\centering
 \includegraphics[scale=0.2]{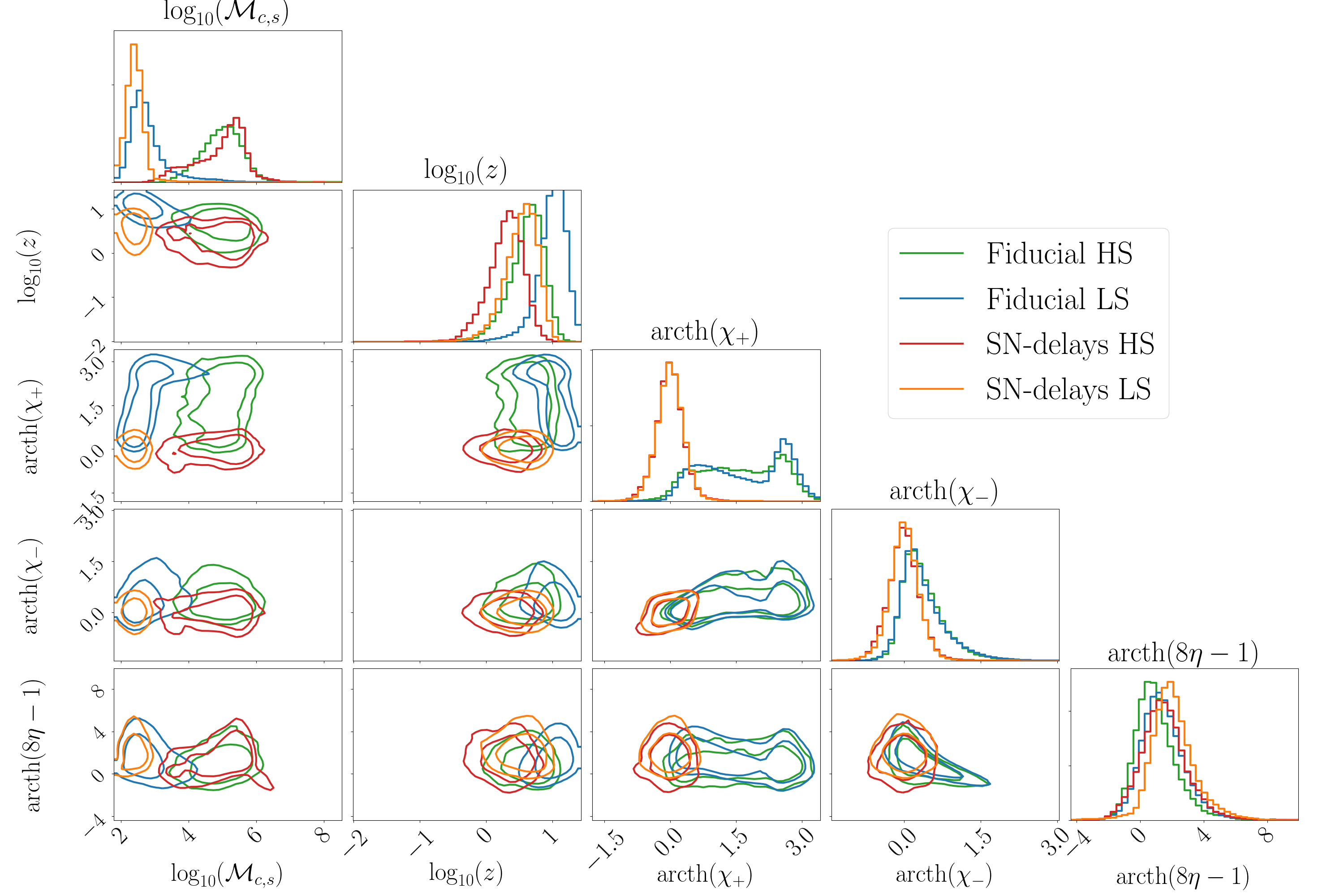}\\
 \caption{Normalized population distributions predicted by our fiducial model and the SN-delays model, both in the LS and HS scenarios. We show the 68\% and 90\% confidence intervals. While the chirp mass distributions in the two models are quite similar, the redshift and effective spin distributions are not. }\label{fig:comp_models}
\end{figure*}
 
When running the simulations, we use only one of the seeding prescriptions. However, the population of MBHs in the Universe is unlikely to be described by any of these ``pure'' models, but rather by a mixture of models. Following~\cite{Sesana:2010wy}, we introduce a mixing fraction $\alpha$ between the LS and HS scenario and define the full (unnormalized) MBHB population distribution to be
\begin{equation}
 N_{\rm pop}(\theta |\alpha)=\alpha N_{\rm pop}(\theta |{\rm LS})+(1-\alpha) N_{\rm pop}(\theta |{\rm HS}).\label{mix}
\end{equation}
In the following, we will denote the normalized population distribution by $p_{\rm pop}(\theta|\alpha)$ and the predicted rate by $R_{\rm ev}$ (in ${\rm yr}^{-1}$), such that $N_{\rm pop}(\theta |\alpha)=R_{\rm ev}(\alpha)p_{\rm pop}(\theta|\alpha)$, with similar definitions for the LS and HS models. The rate for a given value of the mixing fraction is
\begin{equation}
   R_{\rm ev}(\alpha) = \int N_{\rm pop}(\theta |\alpha) {\rm d}\theta = \alpha R_{\rm ev}(LS) + (1-\alpha)R_{\rm ev}(HS),\label{mix:Nev}
\end{equation}
where $R_{\rm ev}(LS) = \int N_{\rm pop}(\theta |\alpha) {\rm d}\theta$ is the rate for the LS model, and similarly for HS.
%obtained through linear interpolation between the value predicted by the LS ($\alpha=1$) and HS ($\alpha=0$) scenarios. %The normalized population distribution is then given by
%
%\begin{equation}
 % p_{\rm pop}(\theta |\alpha)=\frac{\alpha N_{\rm ev}({\rm LS})}{\alpha N_{\rm ev}({\rm LS})+(1-\alpha) N_{\rm ev}({\rm HS})} p_{\rm pop}(\theta |{\rm LS})+\frac{(1-\alpha) N_{\rm ev}({\rm HS})}{\alpha N_{\rm ev}({\rm LS})+(1-\alpha) N_{\rm ev}({\rm HS})} p_{\rm pop}(\theta |{\rm HS}). \label{mix_norm}
%\end{equation}
%

For a given SNR threshold, we denote by $R_{\rm det}(\alpha,{\rm SNR})$ the number of events (per year) above this threshold. In Table \ref{tab:rates} we provide the annual rates for the LS and HS scenarios\footnote{Note that we use a different noise curve and SNR threshold than~\cite{Bonetti:2018tpf,Barausse:2020mdt}, hence the difference in the rates of detectable events.}, as well as the number of detectable events by LISA assuming an SNR threshold of 10, which we use in the remaining of the paper. For comparison, we also give the results for an SNR threshold of 20. The LS scenario predicts more merger events, but many of these have low SNR and are not detectable by LISA. On the contrary, almost all events in the HS scenario are detectable. 
%For an SNR threshold of 20 we actually expect to detect more events in the HS scenario. 

{
\renewcommand{\arraystretch}{1.5}
\begin{table} 
  \begin{center}
   \begin{tabular}{c c c c}
   
%    \cline{3-4}
    
    & & LS & HS \\
    
    \hline
    
    \multicolumn{1}{c}{\multirow{3}{*}{Fiducial}} & $R_{\rm ev}$ (${\rm yr}^{-1}$) &$234.3$ & $23.98$ \\
    
%    \cline{2-4}
    
    \multicolumn{1}{c}{} & $R_{\rm det}(10)$ (${\rm yr}^{-1}$) & $53.01$ & $23.89$  \\
    
%    \cline{2-4}
    
    \multicolumn{1}{c}{} & $R_{\rm det}(20)$ (${\rm yr}^{-1}$) & $29.85$ & $23.67$  \\
    
    \hline \hline
    
     \multicolumn{1}{c}{\multirow{3}{*}{SN-delays}} & $R_{\rm ev}$ (${\rm yr}^{-1}$) & $11.82$ & $5.94$ \\
     
%     \cline{2-4}
     
     \multicolumn{1}{c}{} & $R_{\rm det}(10)$ (${\rm yr}^{-1}$)  &$1.11$ & $5.92$  \\
     
%     \cline{2-4}
     
     \multicolumn{1}{c}{} & $R_{\rm det}(20)$ (${\rm yr}^{-1}$)  &$0.29$ & $5.73$ \\
     
     \hline

   \end{tabular}
   \end{center}
    \caption{Number of events per year $N_{\rm ev}$ and number of detectable events per year with LISA with two different SNR thresholds, $R_{\rm det}(10)$ and $R_{\rm det}(20)$. The LS scenario predicts more events than the HS one, but many of them are not detectable by LISA. Rates in the SN-delays models (bottom) are substantially lower than in our fiducial model (top).}\label{tab:rates}
  \end{table}
  }

In Fig.~\ref{fig:distrib_fid} we show the normalized population distribution for different values of $\alpha$ in a "corner plot"~\cite{corner}. In the lower panel we show only events that have an SNR above 10. We use ``transformed'' parameters (e.g.~$\log_{10} \mathcal{M}_{c,s},\,{\rm arcth}\, \chi_+$) to make the salient features of the distributions more evident. As expected, the HS model predicts binaries with higher masses than the LS model. When mixing between them, we get a double-peaked distribution, whose relative weights depend on the value of $\alpha$. After imposing an SNR cut, lighter events are suppressed, and the relative weights change due to the fact that many LS events are not detectable. The effect of the SNR cut can be clearly seen in the redshift distribution: high-redshift events predicted in the LS scenario are not detectable, and as a consequence the LS and HS redshift distributions after the cut look much more similar. On the contrary, the effective spin distributions are easier to distinguish after imposing the SNR cut. This is because of the correlation between effective spin, redshift and chirp mass, which can be seen in the upper panel. The physical explanation is that the events that survive the SNR cut in the LS scenario tend to be closer
and more massive (both because of the SNR threshold and because the BHs had more time to grow via accretion and mergers). 
Accretion also leads to larger spins for this subset of the population.
Moreover, the presence of gas around binaries tends to align the spins through the Bardeen-Petterson effect, which in turn translates into larger values of the effective spin.
%It is clear from these figures that the (source-frame) chirp mass is the parameter that can better help estimating the value of $\alpha$ from LISA observations. We expect that, to a smaller extent, the redshift and the effective spin could also help. 

%As will be further detailed in section \ref{results}, we are also interested in the compatibility between astrophysical models for LISA data analysis. For this reason, we also consider a more recent class of models, more specifically the SN-delays model of~\cite{Barausse:2020mdt}. This model uses the same seeding prescriptions as our fiducial model, but includes supernovae feedback, which can eject matter around the BHs and quench their growth, and implements additional delays between when dark matter halos merge and MBHBs form in addition to the dynamical friction timescale. 

In Fig.~\ref{fig:comp_models} we compare the normalized population distribution predicted by the SN-delays model to our fiducial model, both in the LS and HS cases, without any SNR threshold. Notice that the chirp mass distributions of the fiducial and SN-delays models are reasonably similar, but the redshift and effective spin ones are very different. The glaring difference in redshift distributions is due to the additional delays included in the SN-delays model, whereas the one in spin distributions is due to supernova feedback, which expels the gas surrounding the BHs in shallow potential wells, resulting in binaries with more isotropic spin orientations and smaller component spin magnitudes.

In Table \ref{tab:rates} we also provide the rates predicted by the SN-delays model. We see that the rates not only differ substantially between the LS and HS scenarios, but also between the fiducial and SN-delays model. A simple way to provide robustness to this rate variation is to introduce an additional parameter into the model, allowing both the mixing fraction $\alpha$ and the total number of events over the observation period $N_\alpha$ to be hyperparameters that we constrain using the observed events. Although we will ultimately marginalize over the number of observations and focus on the mixing parameter, this approach ensures that our inference will be robust as long the model can match the parameter distribution of events, even if the total number of events varies significantly from the semianalytic model predictions.

% shows that it is not a robust prediction of semianalytic models, therefore we will not use rate information when estimating the mixing fraction.
%\eb{Well, if we can't trust rates, why should we trust the rest of the predictions? This can be said better...} \jg{Yes. Is there a parameter, e.g., the fraction of halos seeded with BH seeds, that basically just affects the overall rate that we can argue is less well constrained than other parameters of the model that impact primarily the distribution?}

\section{Hierarchical Bayesian analysis}\label{sec_hba}

Assuming that MBHB events are distributed following the mixing prescription of Eq.~\eqref{mix}, and introducing the overall number of events as an additional parameter characterizing the population, as described in the previous section, the population distribution is described by two hyperparameters, $\alpha$ and $N_\alpha$. By observing many events, we will measure the distribution of MBHB parameters $\theta$ (such as masses, spins and redshifts), and from this we will be able to infer the hyperparameters. Working in a Bayesian framework, our goal is to estimate the posterior distribution of the hyperparameters from a set of observed MBHB events, ${\bf d}$. To do so, we use a similar approach to the ``top-down'' derivation of~\cite{Mandel:2018mve}. We assume that each MBHB event is independently drawn from the population distribution $p_{\rm pop}(\theta|\alpha,N_\alpha)$. Independence is a highly nontrivial assumption for LISA, since the data stream will contain many signals at the same time, from sources of different types, including extreme mass ratio inspirals, Galactic binaries and MBHBs. However, given the expected event rates for LISA sources (see Table \ref{tab:rates}) and the long duration of the LISA mission, these sources are unlikely to have significant overlap with one another. %in the time-frequency plane. 
As a result each source will be sensitive to an independent set of components of the instrumental noise.
This means that it should be reasonable to treat each MBHB observation as independent. 

Under this assumption the probability that, in a certain observation period, a total of $N_t$ events occur in the Universe, with parameters $\pmb{\theta}$, and producing associated strain data, ${\bf d}$, in the detector, is given by
%The joint probability of strain data from all MBHBs, $\{d_k\}$ for $k\in[1,\cdots,N]$, and their associated model parameters, $\theta_k$, is given by
%
\begin{equation} \label{eq:joint_like}
    p({\bf d}, \pmb{\theta},N_t|\alpha,N_\alpha) =  p({\bf d}|\pmb{\theta},N_t)p_\mathrm{pop}(\pmb{\theta},N_t|\alpha,N_\alpha).
\end{equation}
%
%where $N$ is the total number of gravitational wave signals (detectable or not) before thresholding is considered. With our assumption of non-overlapping signals, the 
Assuming that the population of MBHBs is described by a mixture between two independent populations, the second term can be modeled as a Poisson distribution
\begin{align}
    p(\pmb{\theta},N_t|\alpha,N_\alpha)& \propto N_\alpha^{N_t} e^{-N_\alpha} \nonumber \\
    &\prod_{k=1}^{N_t} \left[ f(\alpha) p_\mathrm{pop}(\theta_k|{\rm LS})+ (1-f(\alpha)) p_\mathrm{pop}(\theta_k|{\rm HS})\right],\label{eq:fullpop}
\end{align}
where
\begin{equation}
    f(\alpha) = \frac{\alpha R_{\rm ev}({\rm LS})}{\alpha R_{\rm ev}({\rm LS})+(1-\alpha) R_{\rm ev}({\rm HS})}\label{eq:falphadef}
\end{equation}
is the expected fraction of events in the Universe that come from the LS population.
%$N_{\rm LS}(\alpha)$ and $N_{\rm HS}(\alpha)$ are the number of events coming from the LS and HS populations respectively during the length of the observation. We can write $N_{\rm LS}(\alpha) = \beta N_{\alpha}$, $N_{\rm HS}(\alpha)= (1-\beta) N_{\alpha}$, where $\beta \in [0,1]$ is the fraction of events in the light seed population and $N_{\alpha}$ is the total number of events in the Universe. The fraction $\beta$ is related to the mixing fraction via

Not all the $N_t$ events that occur are detectable. Whether the $k$'th event is detectable is a property of the associated data, $d_k$, only. As shown in~\cite{Mandel:2018mve}, assuming the events are statistically independent, substituting Eq.~\eqref{eq:fullpop} into Eq.~\eqref{eq:joint_like} and marginalizing over the unobserved data yields 
%, then partition the joint likelihood into detected ($i\in[1,\cdots,N_\mathrm{obs}]$) and undetected events, followed by a marginalization over the data, properties, and number of undetected events. This results in 
the following joint likelihood for the detected events:
\begin{align}
        p({\bf d}, \pmb{\theta},N_t|\alpha,N_\alpha) \propto& \exp\left\{-N_{\alpha}(f(\alpha) \Xi({\rm LS}) +(1-f(\alpha)) \Xi({\rm HS})\right\} \nonumber \\
        \times& N_\alpha^{N_{\rm obs}} \prod_{i=1}^{N_\mathrm{obs}} p(d_i|\theta_i)\left(f(\alpha) p_\mathrm{pop}(\theta_i|{\rm LS}) \right . \nonumber \\
       &\left .  +(1-f(\alpha)) p_\mathrm{pop}(\theta_i|{\rm HS})\right),
\label{eq:sepratelike}
\end{align}
where $N_{\rm obs}$ is the number of above threshold events observed and
$\Xi({\rm LS})=R_{\rm det}({\rm LS})/R_{\rm ev}({\rm LS})$ is the fraction of events in the LS population expected to be detectable, which is given by
%normalization factor that accounts for selection effects. 
% The posterior distribution for $\alpha$ is given by Bayes' theorem:
% %
% \begin{equation}
%  p(\alpha|{\bf d})=\frac{ p({\bf d}|\alpha)p(\alpha)}{p({\bf d})}.
% \end{equation}
% For $N_{\rm obs}$ statistically independent signals, the hyperlikelihood can be written as
% %
% \begin{equation}
%  p({\bf d}|\alpha)=\prod_{i=1}^{N_{\rm obs}} p(d_i|\alpha).
% \end{equation}
%This is a crucial consideration, as %We must take into account that 
%not all events will be detectable due to the intrinsic loudness of a given signal and noise fluctuations in the detector. We use the SNR as a detection statistic, and for each possible event in the data $d$ we define
%
%\begin{equation}
%I(d) =
%	\begin{cases}
%		& 1 \; {\rm if} \; {\rm SNR}[d]>{\rm SNR}_{\rm threshold} \\
%		& 0 \; {\rm otherwise.}
%	\end{cases} \label{mbbhs:det_stat}
%\end{equation}

% The individual hyperlikelihood for an event can then be written as
% %
% \begin{equation}
%  p(d|\alpha)=\frac{1}{\Xi(\alpha)} \int {\rm d} \theta  \ p(d|\theta)p_{\rm pop}(\theta|\alpha)I(d).
% \end{equation}
%The normalisation factor $\Xi(\alpha)$ %(sometimes called \textit{selection function}) 
%is given by
%
\begin{align}
 \Xi({\rm LS})&=\int {\rm d \theta} \  p_{\rm pop}(\theta|{\rm LS}) \int_{d,{\rm detectable}} {\rm d}d  \ p(d|\theta) \nonumber \\
 &=\int {\rm d \theta} \  p_{\rm pop}(\theta|{\rm LS}) p_{\rm det}(\theta),
\end{align}
where the last equality defines $p_{\rm det}(\theta)$, the probability of detecting an event with parameters $\theta$. The quantity $\Xi({\rm HS})$ is defined in an analogous way for the HS population. In this work we use the SNR to quantify detectability and assume that an event, $d$, is detectable if SNR$[d] > $SNR$_{\rm threshold}$. Since we work in the zero-noise approximation, we evaluate this using the optimal SNR to determine the detectability of each source. The selection function, $\Xi({\rm LS})$, is equal to the fraction of events in the population that have SNR above the threshold. 

The final form of the posterior distribution on $\alpha$ and $N_\alpha$ is obtained by marginalization over the parameters of the individual events, ${\pmb \theta}$, in Eq.~\eqref{eq:rate_marglike} and using Bayes' theorem. After some rearrangement we obtain
%, and a reshuffling of terms according to Bayes' Theorem, such that
%
\begin{align}
    p(\alpha,N_\alpha|{\bf d}) &= \frac{ p({\bf d}|\alpha,N_\alpha)p(\alpha,N_\alpha)}{p({\bf d})} \nonumber\\
    &\propto \frac{p(\alpha,N_\alpha)\prod_{i=1}^{N_\mathrm{obs}} p(d_i)}{p({\bf d})} N_\alpha^{N_{\rm obs}} \exp[-N_\alpha \Xi(\alpha)] \nonumber \\
    &\hspace{0.5cm}\times\prod_{i=1}^{N_\mathrm{obs}} \int d\theta_i \frac{p(\theta_i|d_i)p_\mathrm{pop}(\theta_i|\alpha)}{p_i(\theta_i)},
\end{align}
in which $p(\theta_i|d_i)=p(d_i|\theta_i) p_i(\theta_i)/p(d_i)$,  $p_i(\theta_i)$ denotes the prior used to obtain some posterior samples in an initial analysis of event-$i$, and we have introduced
%For model~\eqref{mix} it can be related to the selection functions of the pure models via
%
\begin{align}
 \Xi(\alpha)&=f(\alpha)  \ \Xi({\rm LS})+(1-f(\alpha)) \Xi({\rm HS})\label{selection} \\
    p_{\rm pop}(\theta|\alpha)&=f(\alpha) p_\mathrm{pop}(\theta|{\rm LS})+ (1-f(\alpha))p_\mathrm{pop}(\theta|{\rm HS}).
\end{align}
In an analysis of LISA data we would construct this posterior on both hyperparameters. However, the parameter of most interest is the mixing fraction $\alpha$, and so we will focus on this here. We proceed by marginalizing over the rate parameter, $N_\alpha$. We first specify that the hyperprior is separable, $p(\alpha,N_\alpha)=p(\alpha)p(N_\alpha)$, and then assume a scale-invariant prior on the rate, $p(N_{\alpha})\propto 1/N_{\alpha}$. The scale-invariant $1/N_{\alpha}$ prior is natural when the order of magnitude of the rate is uncertain, as is the case here. After this marginalization we obtain
%, a effectively states that the number of events conveys no information about the properties of the population. While this is not always true it should be a conservative approximation. We note that it is the intrinsic population distribution, $p_{\rm pop}(\theta|\alpha)$, that appears above, not the population distribution for detectable events. Selection effects are entirely contained in the selection function $\Xi(\alpha)$

%Equation~\eqref{eq:sepratelike} accounts for both the number and parameter distribution of detected events. However, given the large disparity in rate estimates between different models (see table \ref{tab:rates}), information from the number of events is likely to be unreliable. If we marginalize over the unknown rate using the prior , we obtain
%
%\begin{equation} \label{eq:rate_marglike}
%    p({\bf d}, \pmb{\theta}|\alpha) \propto \prod_{i=1}^{N_\mathrm{obs}} \frac{p(d_i|\theta_i) p_{\rm pop}(\theta|\alpha)}{\Xi(\alpha)},
%\end{equation}
%, and a reshuffling of terms according to Bayes' Theorem, such that
%
\begin{align}
    p(\alpha|{\bf d}) &= \frac{ p({\bf d}|\alpha)p(\alpha)}{p({\bf d})} \nonumber\\
    &\propto \frac{p(\alpha)\prod_{i=1}^{N_\mathrm{obs}} p(d_i)}{p({\bf d})} \prod_{i=1}^{N_\mathrm{obs}} \int d\theta_i \frac{p(\theta_i|d_i)p_\mathrm{pop}(\theta_i|\alpha)}{p_i(\theta_i)\Xi(\alpha)}. \label{eq:rate_marglike}
\end{align}
 If $N_i$ posterior samples have been obtained for event $i$ using the reference prior $p_i(\theta_i)$, these can be used to obtain a Monte Carlo approximation to the integrals in the preceding equation
%Usually, when performing parameter estimation on an event, we collect samples from the posterior distribution. We can thus use %Bayes' theorem to express the events' likelihoods, $p(h_i|\theta_i)$, in terms of their posteriors, $p(\theta_i|h_i)$, and use a 
%Monte Carlo averaging to substitute the integral in the previous equation by a sum over the $N_i$ collected samples $\{\theta_j\}_i$ for event-$i$. This leads to the following form of the posterior distribution on the mixing fraction that we use here in our injections and analyses:
%
\begin{align}
    p(\alpha|{\bf d})&=\prod_{i=1}^{N_{\rm obs}} \left [ \frac{1}{N_i} \sum_{j=1}^{N_i}\frac{p_{\rm pop}(\theta_{i,j}|\alpha)}{p_i(\theta_{i,j}) \Xi(\alpha)} \right ]
    %_{\theta_{i,j} \sim p(\theta|d_i)} 
    \nonumber \\
    & \times p(\alpha) \frac{\prod_{i=1}^{N_{\rm obs}}p(d_i)}{p({\bf d})} ,\label{hba}
\end{align}
where $\theta_{i,j}$ is the parameter vector for the $j$'th sample for source $i$. The individual event and overall evidences, $p(d_i)$ and $p({\bf d})$, are useful for model selection but merely enter as a normalization constant when the interest is on parameter estimation, as here.
%In the above equation, the $p_i(\theta)$ are the priors used to perform parameter estimation on each single event, and the $p(d_i)$ are the events' evidence. Since w
Therefore, we discard all evidence terms from our analysis. For the prior on $\alpha$, we take a flat distribution in $[0,1]$. %Note that the population distribution $p_{\rm pop}(\theta|\alpha)$ is the intrinsic one, i.e.~without applying any SNR cut, and that selection effects are entirely contained in the selection function $\Xi(\alpha)$.  
%The rates information could be included in this formalism by assuming $N_{\rm obs}$ to follow a Poisson distribution of mean $N_{\rm det}(\alpha,{\rm SNR})$, and including this term on the right-hand side of Eq.~\eqref{hba}. Given the large disparity in rates' estimates between different models (see Table~\ref{tab:rates}), we will not use this information. 

We note that the quantity $f(\alpha)$ is directly interpretable as the fraction of events in the Universe that are drawn from the LS model, while the mixing fraction $\alpha$, as we have defined it, is not. However, these are related by the simple transformation given in Eq.~\eqref{eq:falphadef}, and so the posterior for $f(\alpha)$ can readily be derived from that for $\alpha$ and vice versa.

%The parameter $\beta$ has a slightly different interpretation to $\alpha$, as the latter is affected by the difference in the predicted rates for the two models, but it is easy to transform between the two. While we only consider the $\alpha$ parameterisation here, Eq.~\eqref{eq:sepratelike} could also be marginalised for different kinds of rate priors, which would yield alternative forms for the likelihood. \jg{If we are arguing that model rate predictions are unreliable then $\beta$ might be a better parameter as it is directly the fraction of events from the LS population. While this is just a transformation maybe it is too late to consider this now?} 
%
% By definition, all observed events have $I(d_i)=1$, thus
% %
% \begin{equation}
%   p({\bf d}|\alpha)=\prod_{i=1}^{N_{\rm obs}} \frac{\int {\rm d}\theta \  p(d_i|\theta)p_{\rm pop}(\theta|\alpha)}{\Xi(\alpha)} \label{likelihood_hba}.
% \end{equation}

\begin{figure*}[hbtp!]
% \centering
%\subfigure[$\alpha_0=0.2$.]{
%  \centering
  \includegraphics[scale=0.09]{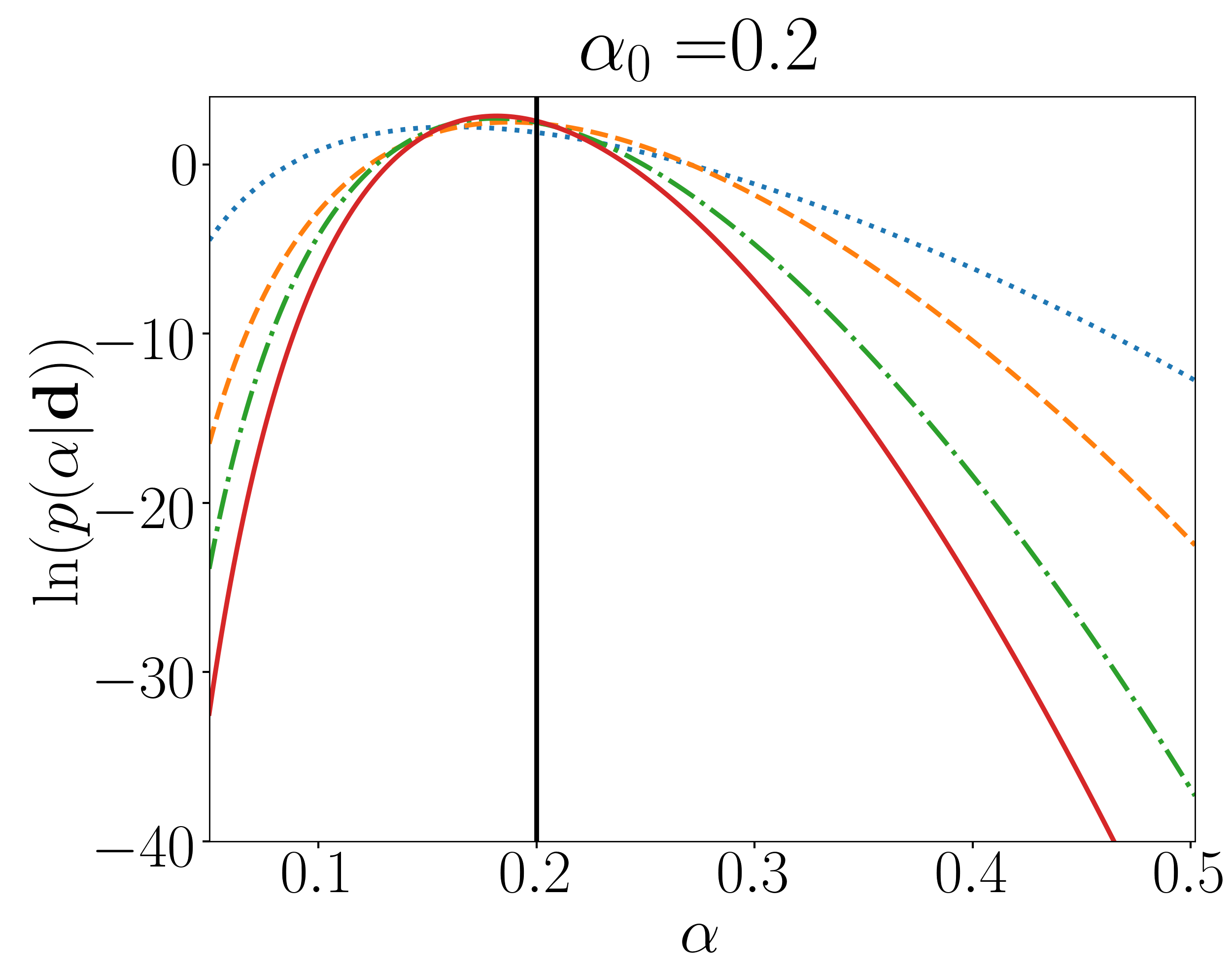}
%    }
%\centering
%\subfigure[$\alpha_0=0.5$.]{
%    \centering
    \includegraphics[scale=0.09]{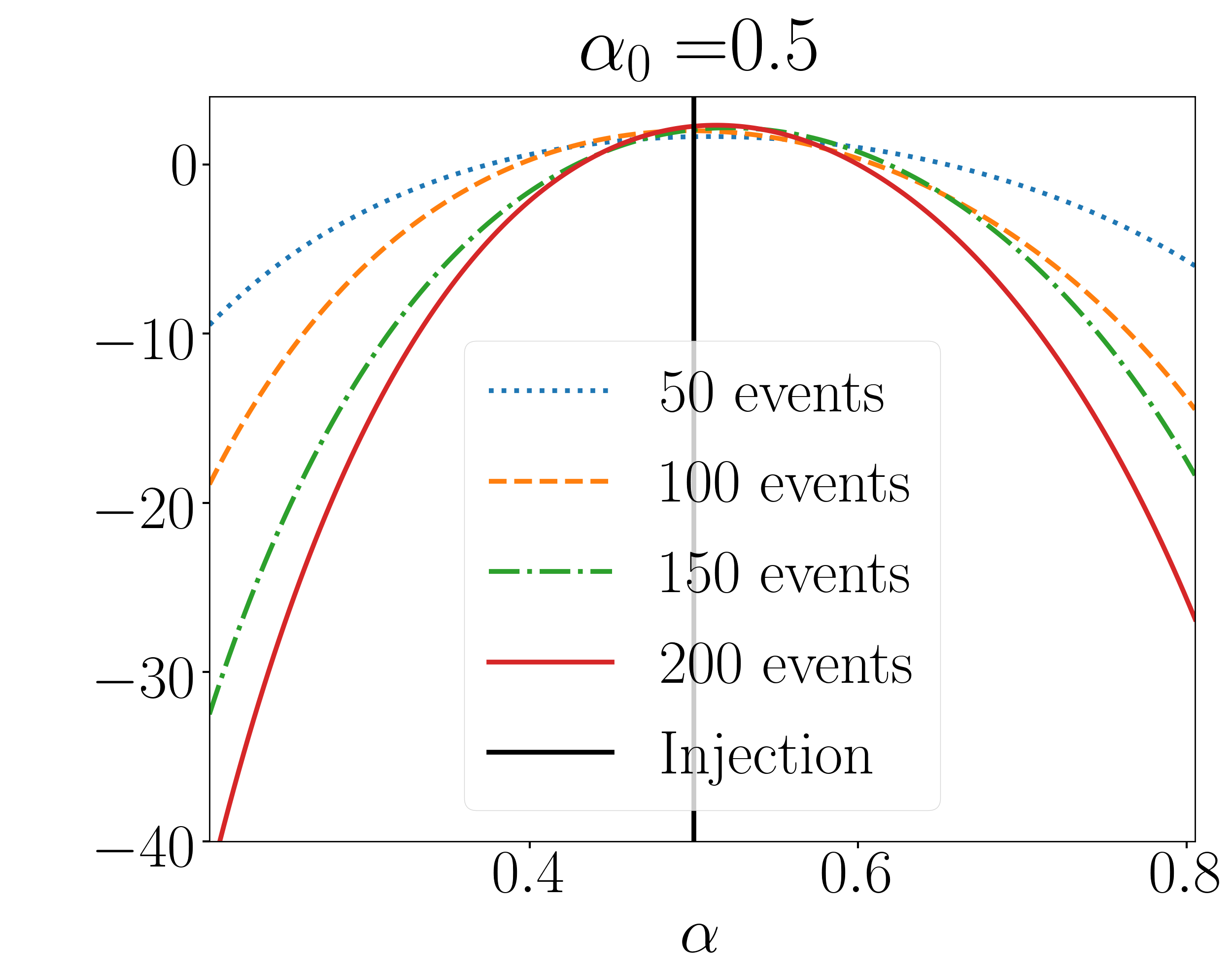}
%   }
%   \centering
%   \subfigure[$\alpha_0=0.8$.]{
%    \centering
    \includegraphics[scale=0.09]{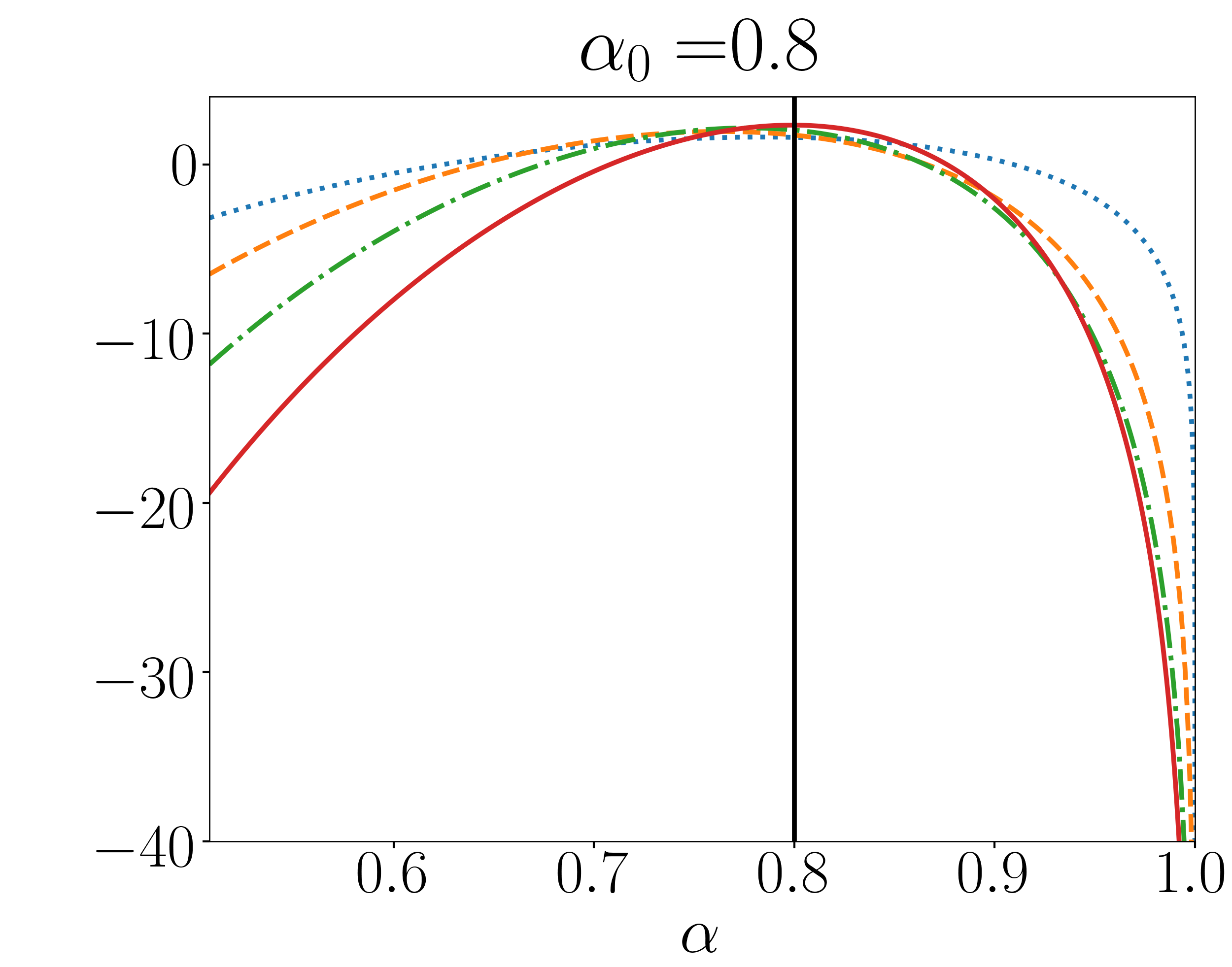}
%   }
%   \centering
   \caption{Posterior distribution on $\alpha$ for observation sets with an increasing number of observed events, generated using different values of the mixing fraction $\alpha_0$: $\alpha_0=0.2$ (left), $\alpha_0=0.5$ (middle) and $\alpha_0=0.8$ (right). The posteriors peak near the true value and become narrower as we increase the number of events.}\label{fig:loglikes}   
 \end{figure*}

%Once we estimate  from an observed data set
After inferring a posterior distribution on $\alpha$, we can construct the posterior predictive distribution (PPD) for the parameters of future observed events
\begin{equation}
 {\rm PPD}(\theta|{\bf d})=\int {\rm d} \alpha  \ p_{\rm pop}(\theta|\alpha) p(\alpha|{\bf d}). \label{ppd} 
\end{equation}
%
%This quantity encodes our prediction on the distribution of population observables ($\theta$) given an observed data set. 
When performing simulations, comparing the PPD with the population distribution used to generate the data provides a guide to the quality of the inference.

\section{Estimating the probability density function}\label{kde}

From Eq.~\eqref{hba}, we can see that the hierarchical Bayesian analysis requires being able to evaluate the probability density function of the population distribution. However, semianalytic models only provide samples from the population distribution, not the analytic probability density function. In this work, we use a kernel density estimator (KDE)~\cite{10.1214/aoms/1177704472,10.1214/aoms/1177728190} to approximate the population probability density function from the samples. More specifically, we use the Gaussian KDE implementation of \texttt{scipy}~\cite{2020SciPy-NMeth}. In Appendix \ref{app:kde}, we provide additional details on how the KDE is computed.

The required accuracy on the estimation of the probability density function increases with the number of observed events. The accuracy of the KDE is limited by the number of simulation points at our disposal, in particular for the HS variant of our fiducial astrophysical model ($\sim 2500 $ points). This leads to a systematic error, which dominates over statistical errors when increasing the number of observed events, and leads to systematic biases in the hierarchical Bayesian analysis. Similarly, from Eq.~\eqref{hba} it can be seen that the error on $\ln(p(\alpha|{\bf d}))$ due to a misevaluation of the selection function increases linearly with the number of observed events. In our case, the accuracy to which the selection function is computed depends on the accuracy of the selection function for the LS and HS models: cf. Eq.~\eqref{selection}. In Appendix~\ref{app:selection}, we show that using too few points to compute these terms also leads to systematic biases. To mitigate these issues, we make an approximation: we take the probability density function computed from the KDE to be the ``true'' probability density function of our fiducial astrophysical model, and use it to generate mock data. By doing this, the data generation process is fully consistent with the probability density function used in the hierarchical Bayesian analysis, avoiding systematic biases. We compute the selection function for the LS and HS variants of our fiducial astrophysical model by generating many ($\sim 10^6$) events from the KDE and computing the fraction of detectable events. We then use Eq.~\eqref{selection} to evaluate the selection function for any value of $\alpha$. This approximation should be seen as the limit where we have enough simulation points to build very accurate KDEs and compute the selection function to high precision. In Appendix~\ref{app:modelcomp} we compare the population distribution of the LS and HS variants of the fiducial astrophysical model computed from numerical simulations to the one obtained from the KDEs, computed as described in Appendix \ref{app:kde}. Note that, when building the KDE that will serve as our fiducial astrophysical model, we use ${\rm arcth}\, \chi_{1,2}$ instead of ${\rm arcth}\, \chi_{+,-}$ to make sure that the spins are in the physically allowed range. The distributions are overall in very good agreement, so we expect that our results should not depend much on this approximation. 

\section{Results}\label{results}

We start by testing our pipeline in the limit where the parameters of the source are perfectly measured by LISA,
%, i.e.~$p(d_i|\theta)=\delta(\theta-\theta_i)$ in Eq.~\eqref{likelihood_hba}, \eb{commented out - fix broken ref}
and we perform two experiments. In the first one (Sec.~\ref{sec:results_a}) we generate mock observation sets using the predictions of our fiducial astrophysical model, as computed from the KDE, and use this same model in the hierarchical Bayesian analysis. %(the $p_{\rm pop}(\theta|\alpha)$ in Eq.~\eqref{hba}).
In the second experiment (Sec.~\ref{sec:results_b}) we use the SN-delays model to generate mock observation sets, but still use our fiducial astrophysical model in the hierarchical Bayesian analysis. The goal of this second experiment is to test if we could still draw meaningful conclusions if the population of MBHBs in the Universe were different from the one used in the data analysis pipeline. In Sec.~\ref{sec:results_c} we discuss the impact of measurement errors in the analysis. In all cases we use an SNR threshold of 10 to define detectability of a source.

\subsection{Model-consistent inference}
\label{sec:results_a}
 
We start by investigating how the inference on $\alpha$ improves with the number of observed events. Although we do not use information on the rates in the inference, we make sure that the number of events in the datasets is realistic for a LISA mission duration of four to ten years, given the predicted rates (see Table \ref{tab:rates}). In Fig.~\ref{fig:loglikes}, we plot the log-posterior on $\alpha$ for observation sets with an increasing number of observed events. In the left panel, the dataset was generated with a mixing fraction $\alpha_0=0.2$ between the LS and HS variants of our fiducial astrophysical model, in the middle panel with $\alpha_0=0.5$, and in the right panel with $\alpha_0=0.8$. The posteriors peak near the true value and become narrower as we increase the number of events. We observe a sharp drop in the posterior close to the extremal values. This is because as $\alpha\to 0$ ($\alpha\to 1$) the resulting population is no longer compatible with the lightest (heaviest) events. Moreover, due to our choice of mixing prescription in Eq.~\eqref{mix} and to the higher event rate of the LS variant, the population distribution varies faster for small values of $\alpha$, so the posterior is narrower for $\alpha_0\simeq 0$ than for $\alpha_0\simeq 1$. 

 \begin{figure}[h!]
\centering
 \includegraphics[scale=0.1]{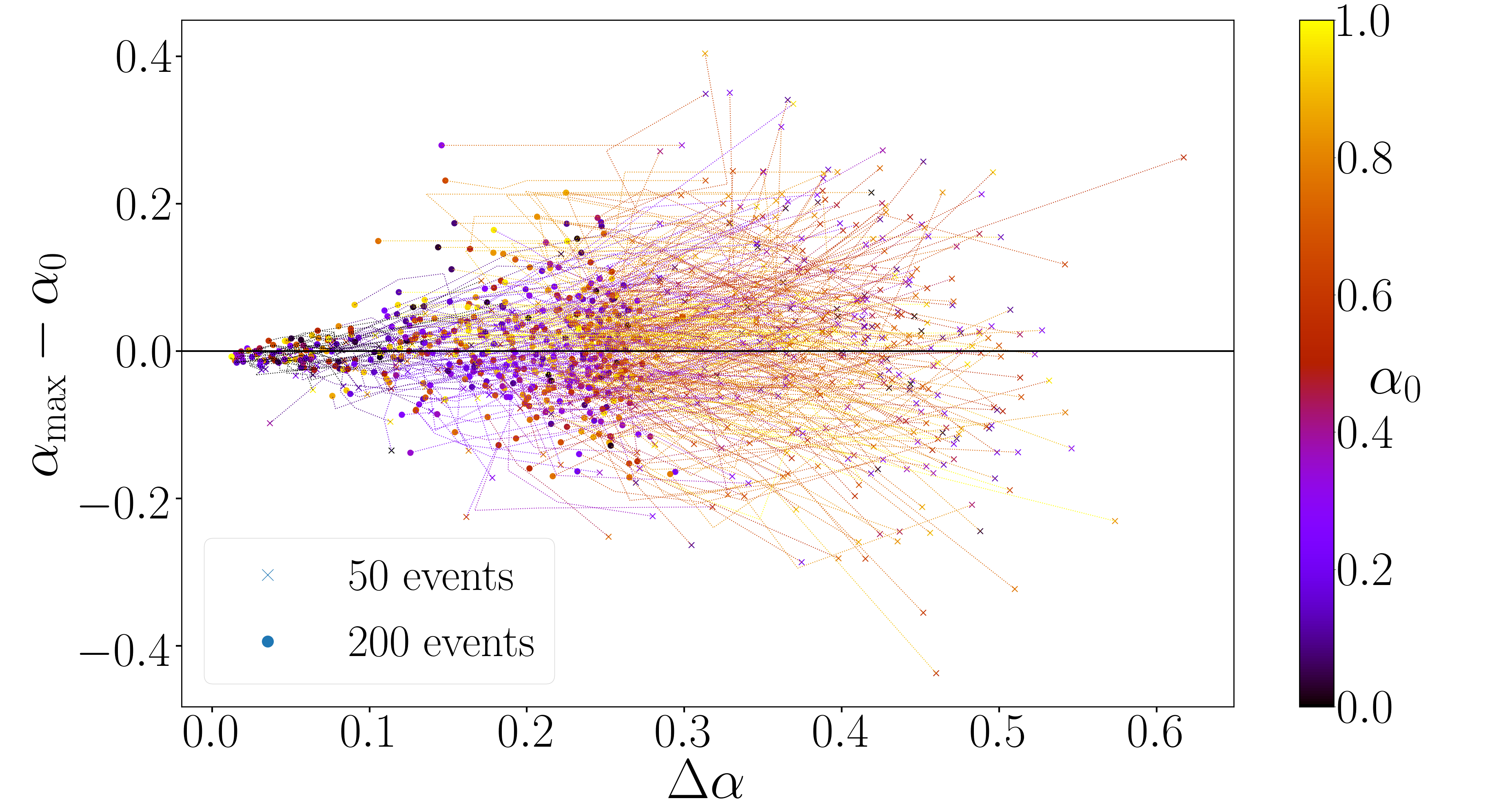}\\
 \centering
 \caption[Evolution of the shift and the error on $\alpha$ (90\% confidence interval) with the number of observed events.]{Evolution of the shift and the error on $\alpha$ (90\% confidence interval) with the number of observed events. We consider two sets of observations, with 50 events (crosses) and 200 events (dots). The color scale indicates the value of $\alpha_0$. As expected, they tend do decrease as we observe more events. The fact that the points are equally distributed on both sides of the $\alpha_{\rm max}=\alpha_0$ line indicates that there is little systematic bias in our analysis.}\label{fig:bias_error}
\end{figure}

\begin{figure*}[hbtp!]
 \centering
\subfigure[ \ 50 events.]{
    \centering \includegraphics[scale=0.09]{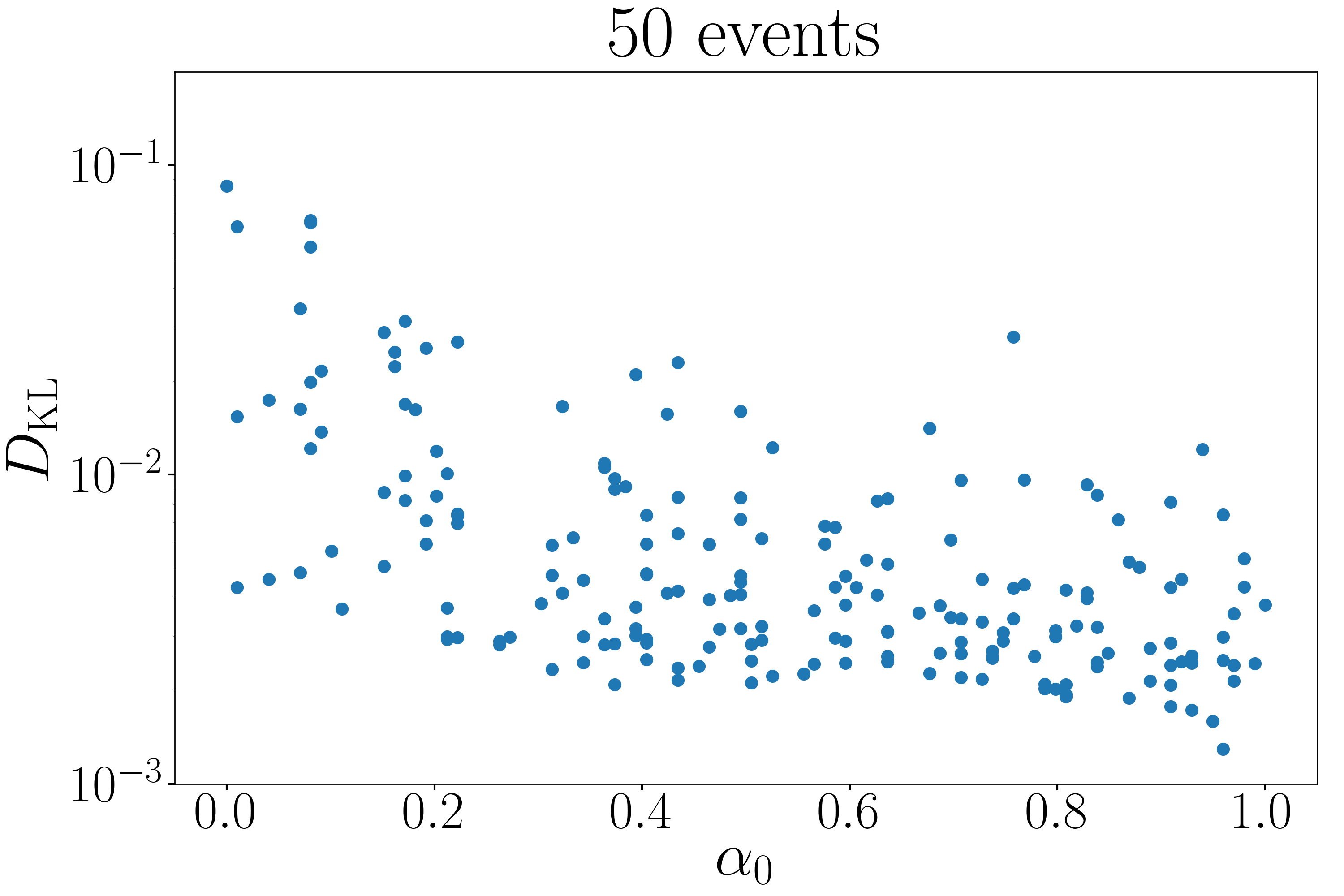}
    }
\centering
\subfigure[ \ 200 events.]{
    \centering \includegraphics[scale=0.09]{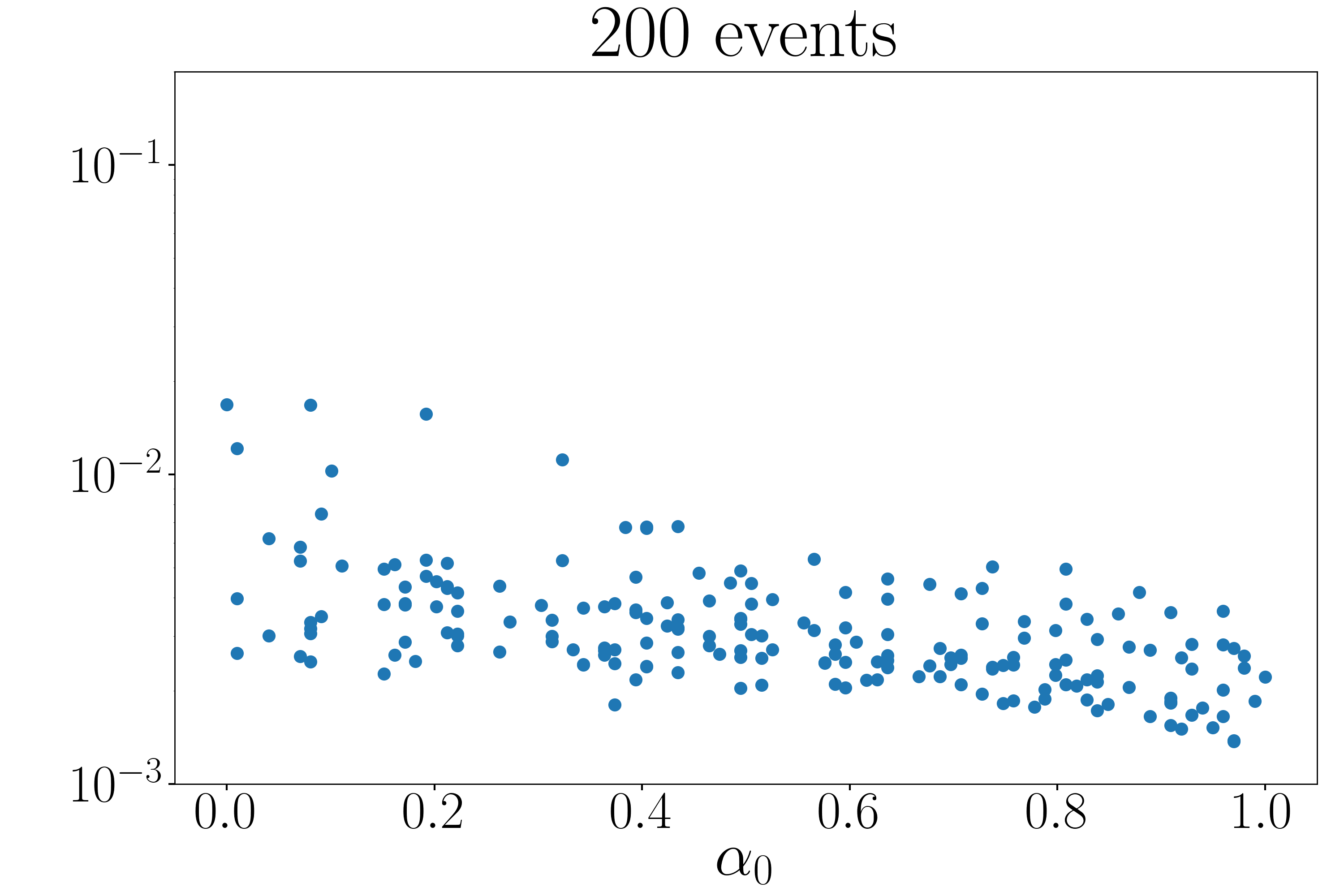}
   }
   \centering
   \caption{Kullback-Leibler divergence between the PPD and the population distribution for different observation sets generated with different values of $\alpha_0$. On the left (right) panel the observation sets contain 50 (200) observed events. The smaller the KL divergence, the better our inference of the population distribution. Increasing the number of events tends to improve the inference, as expected.}\label{fig:ppd_kl}   
 \end{figure*}
 
 \begin{figure*}
 \centering
\subfigure[Worst case.]{
    \centering \includegraphics[scale=0.22]{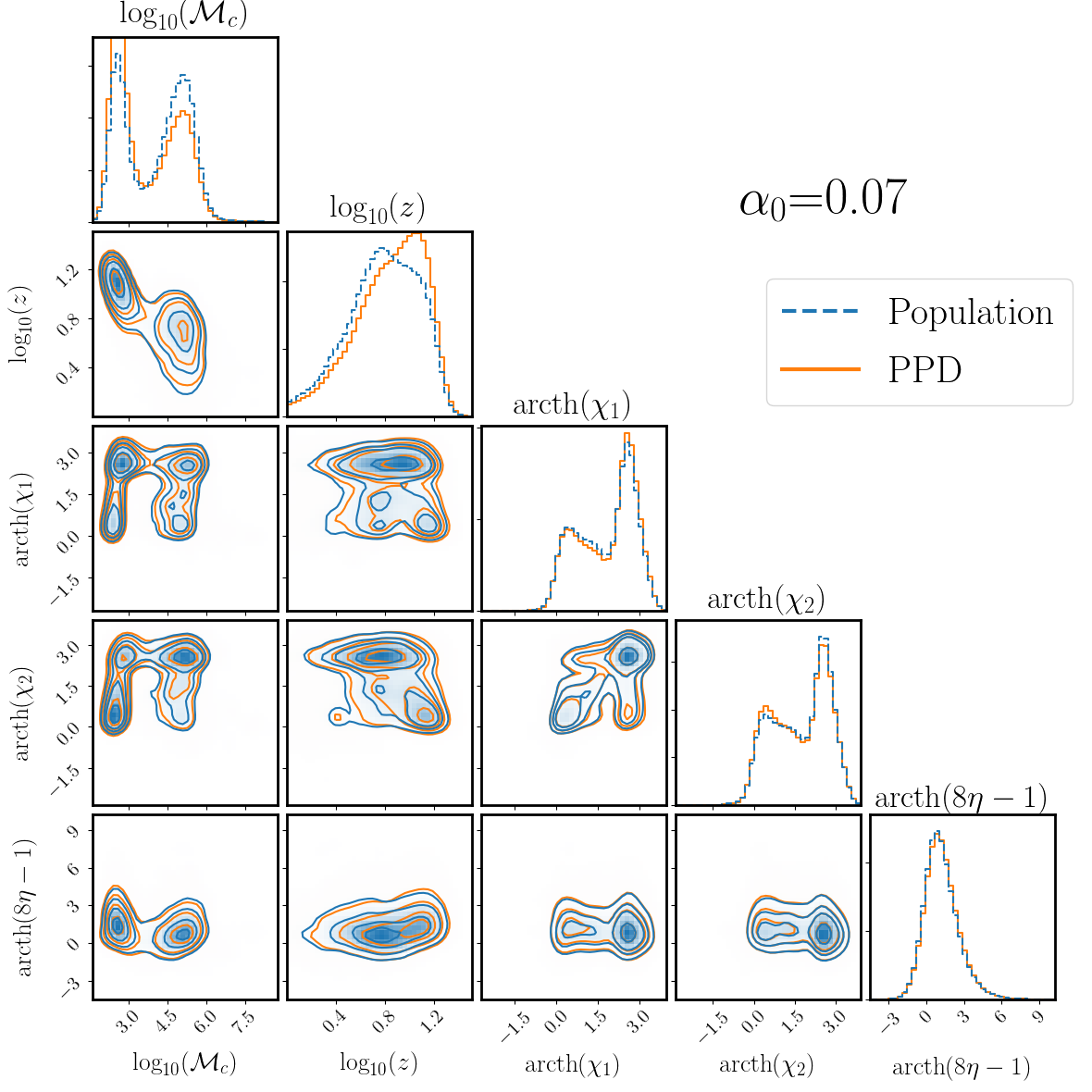}
    }
\centering
\subfigure[Mid-range case.]{
    \centering \includegraphics[scale=0.22]{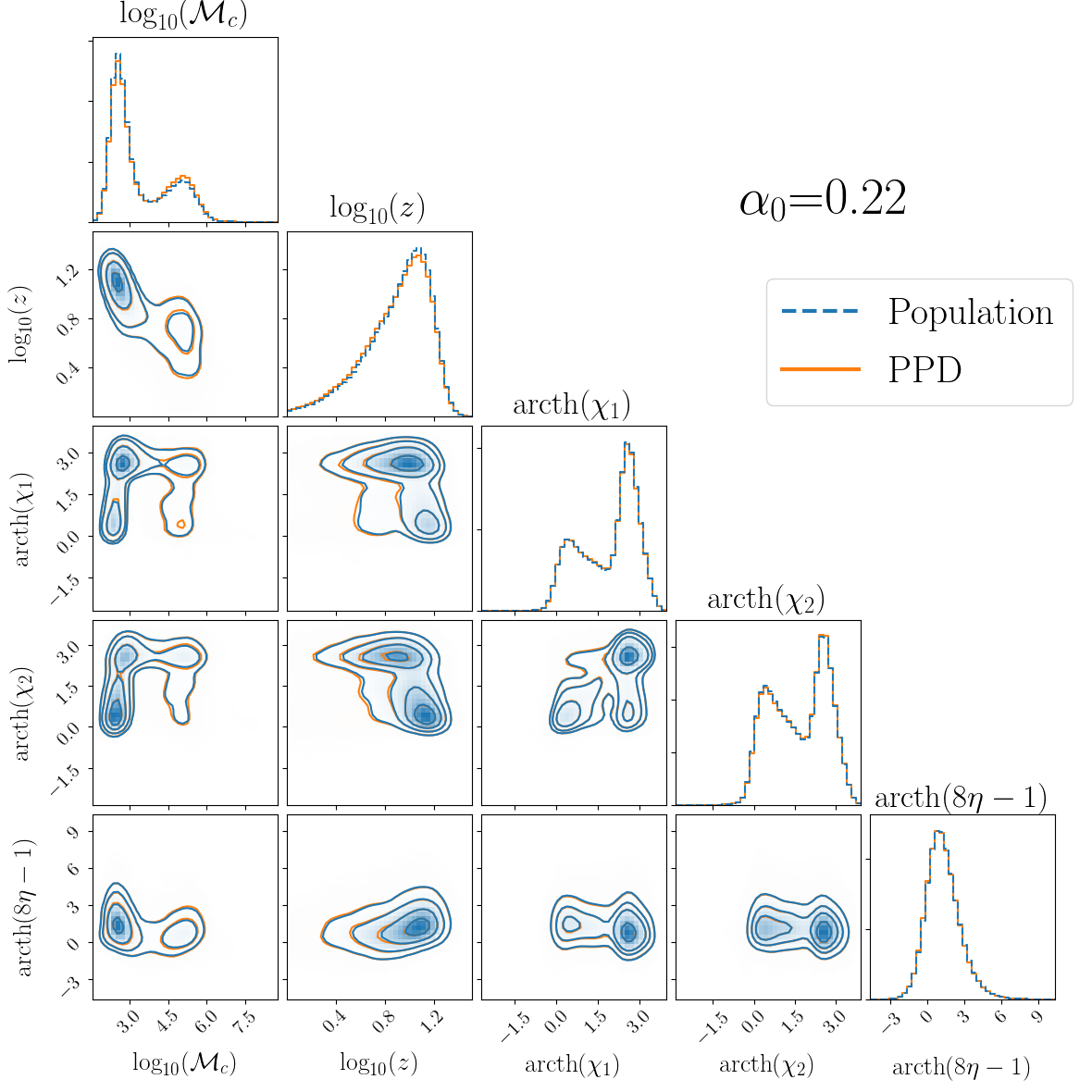}
   }
   \centering
   \centering
\subfigure[Mid-range case.]{
    \centering \includegraphics[scale=0.22]{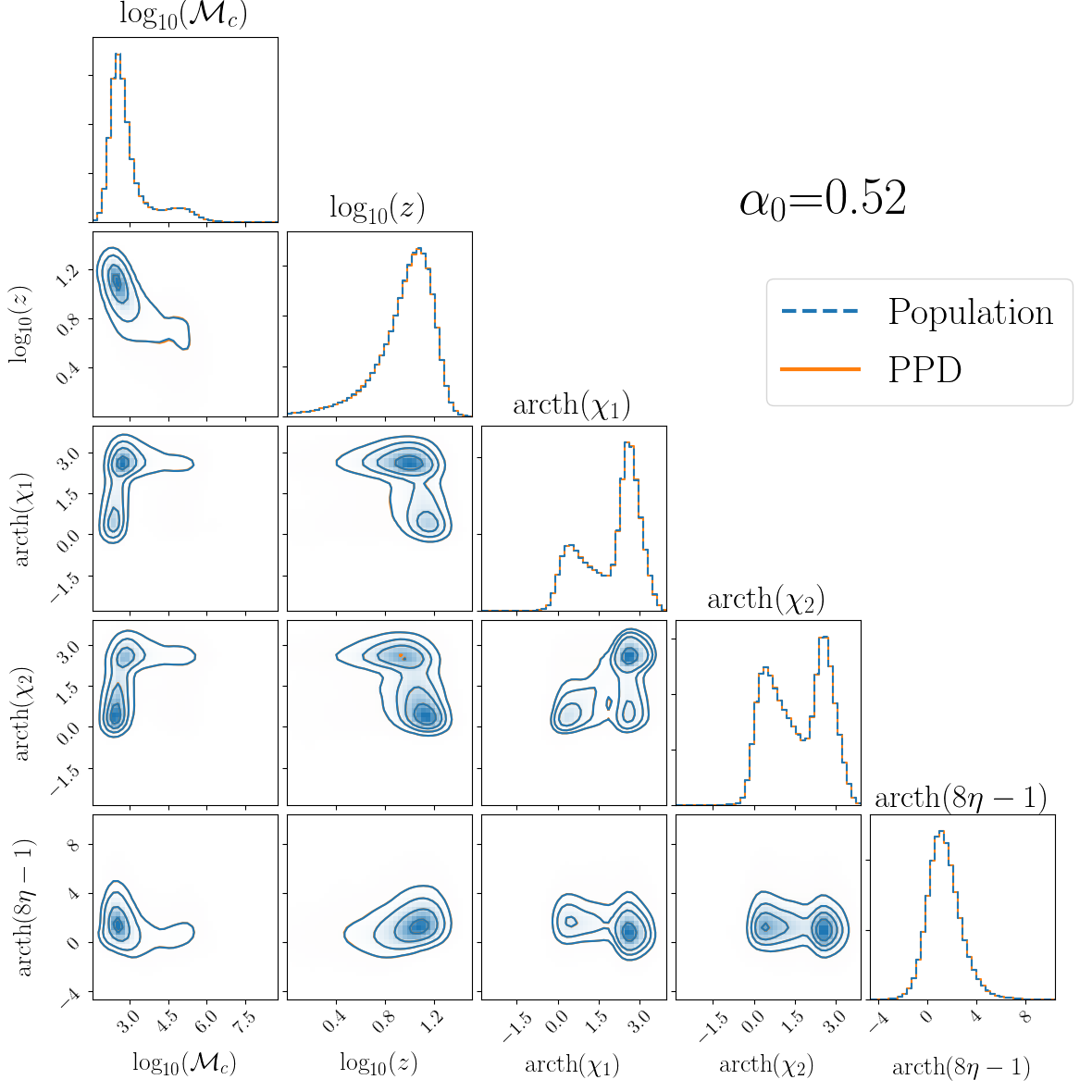}
   }
   \centering
   \centering
\subfigure[Best case.]{
    \centering \includegraphics[scale=0.22]{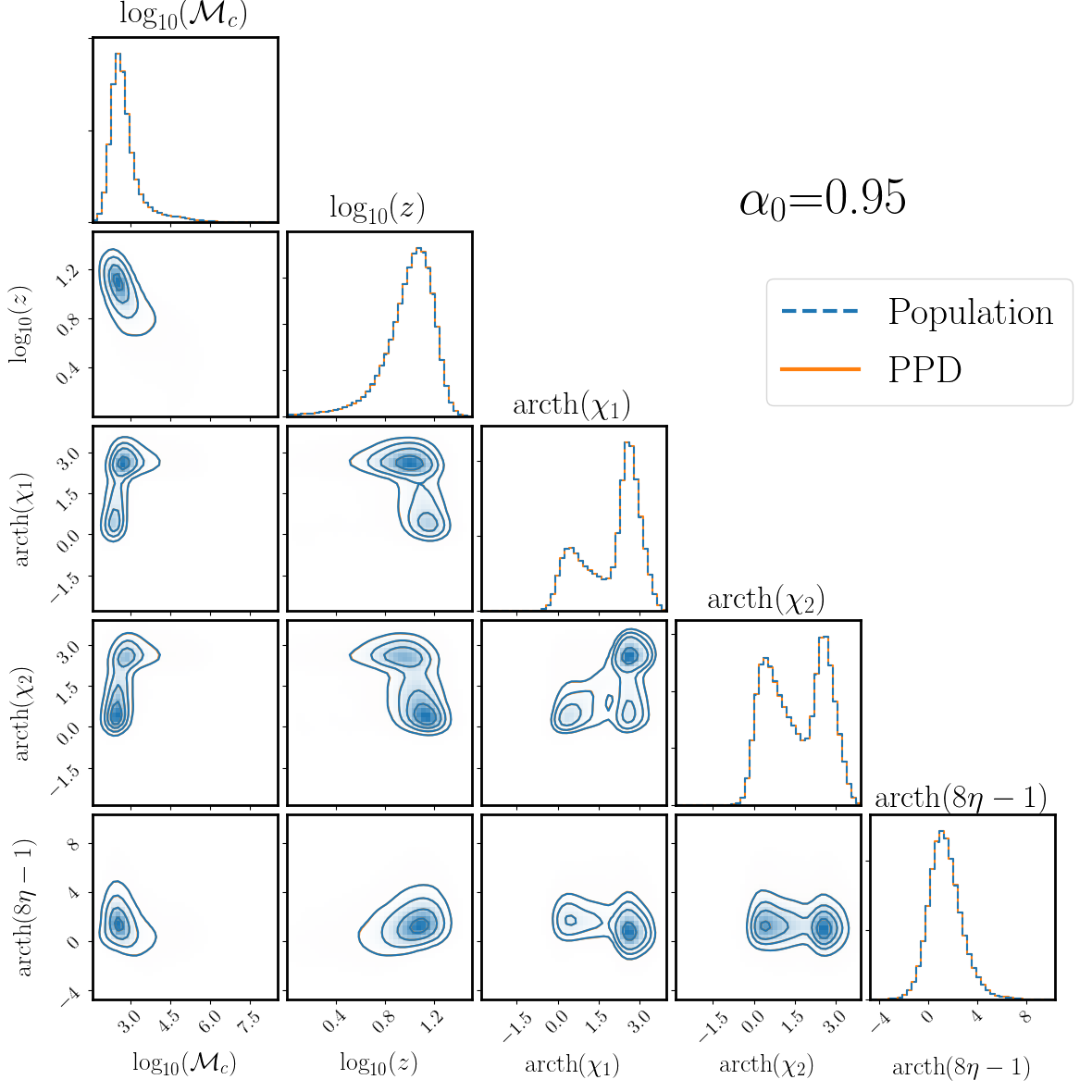}
   }
   \centering
   \caption{Population distribution and PPD for four sets of observations generated with different values of $\alpha_0$. Each observation set contains 100 events. On the upper-left and lower-right panels we show the cases that yield the largest and smallest values of the KL divergence among the cases shown in Fig.~\ref{fig:ppd_kl}. The other two panels show cases yielding mid-range values of the KL divergence. }\label{fig:ppds}   
 \end{figure*}

\begin{figure*}[hbtp!]
 \centering
%\subfigure[Detectable population.]{
    \centering \includegraphics[scale=0.22]{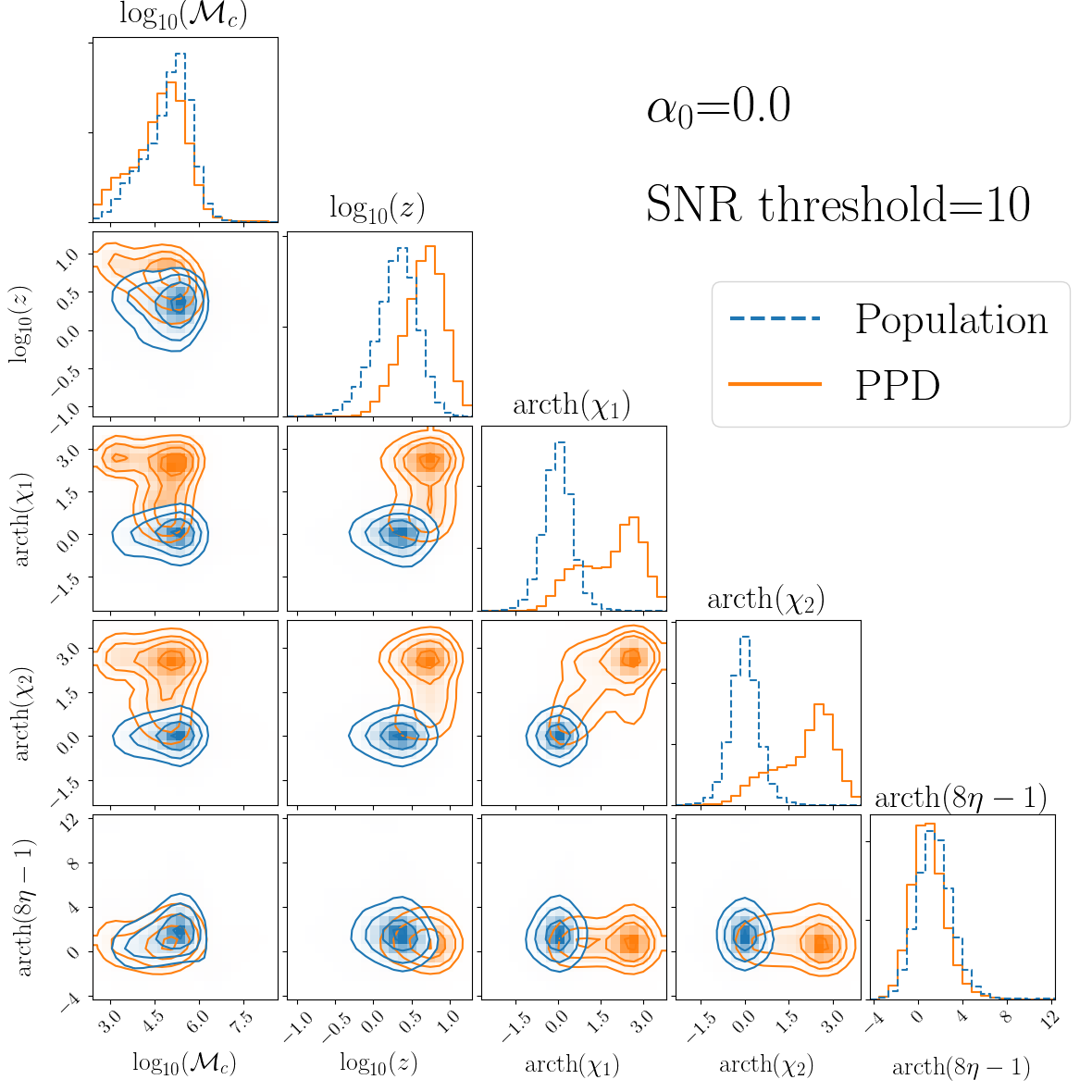}
%    }
\centering
%\subfigure[Intrinsic population.]{
    \centering \includegraphics[scale=0.22]{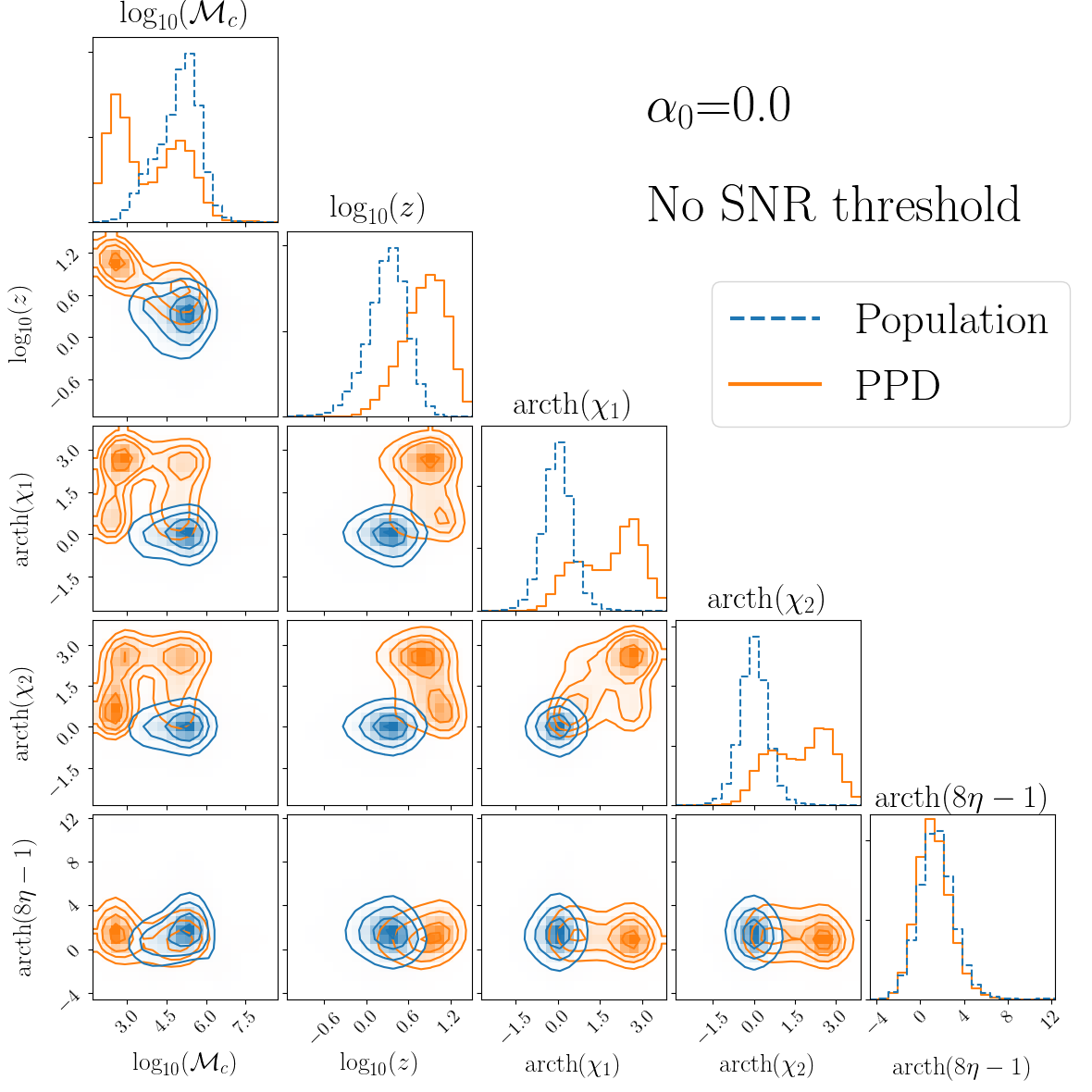}
%   }
   \centering
   \centering
%\subfigure[Detectable population.]{
    \centering \includegraphics[scale=0.22]{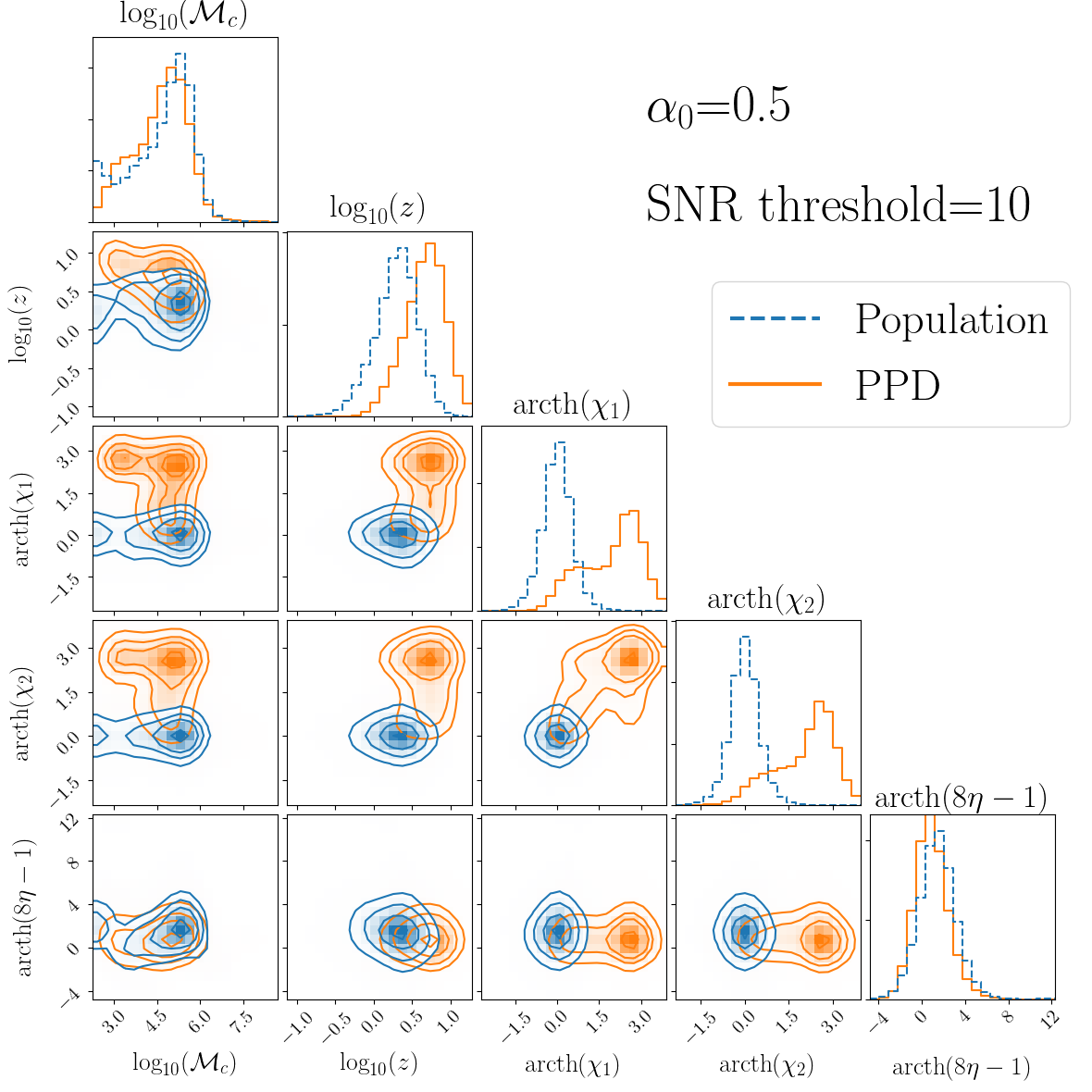}
%    }
\centering
%\subfigure[Intrinsic population.]{
    \centering \includegraphics[scale=0.22]{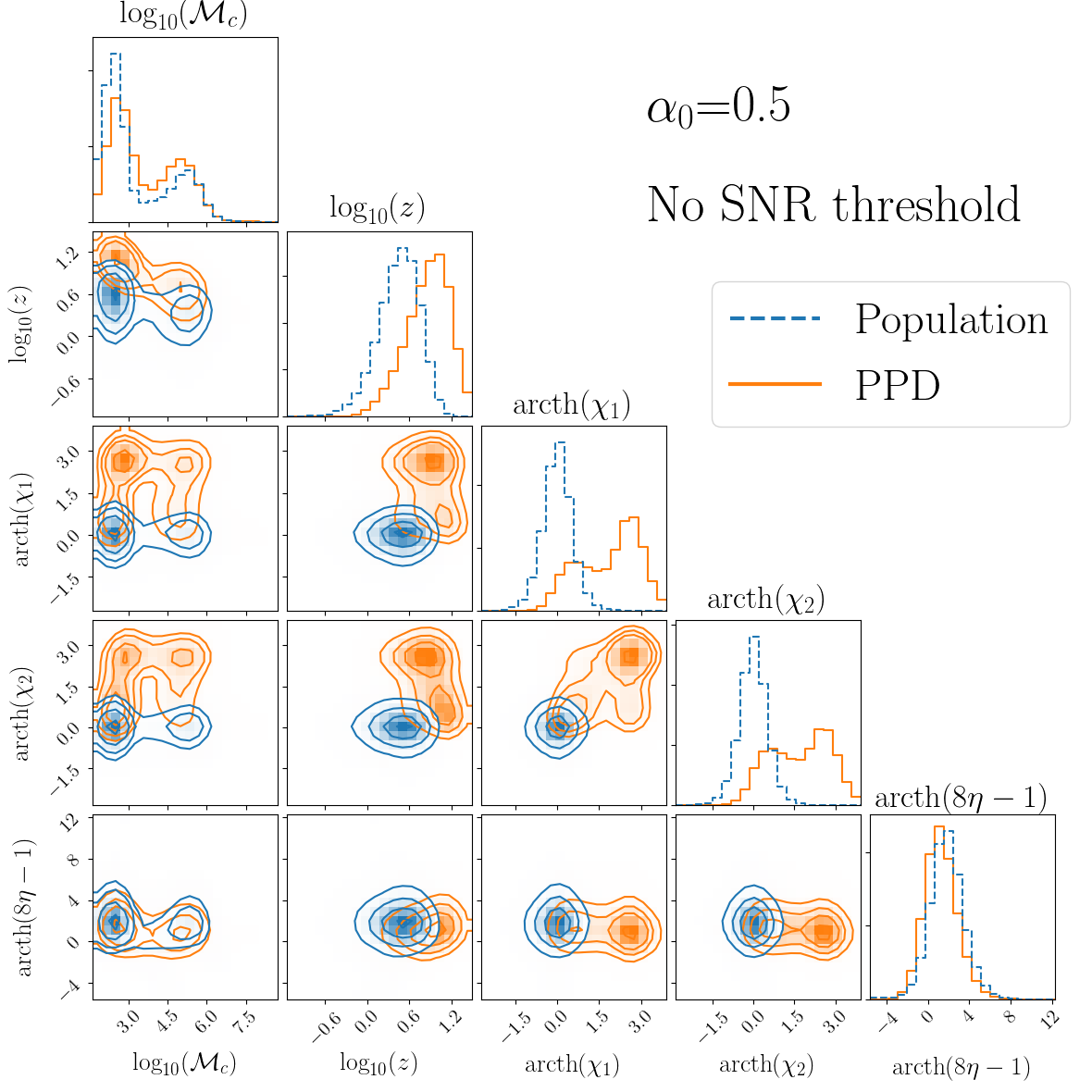}
%   }
   \centering
    \centering
%\subfigure[Detectable population.]{
    \centering \includegraphics[scale=0.22]{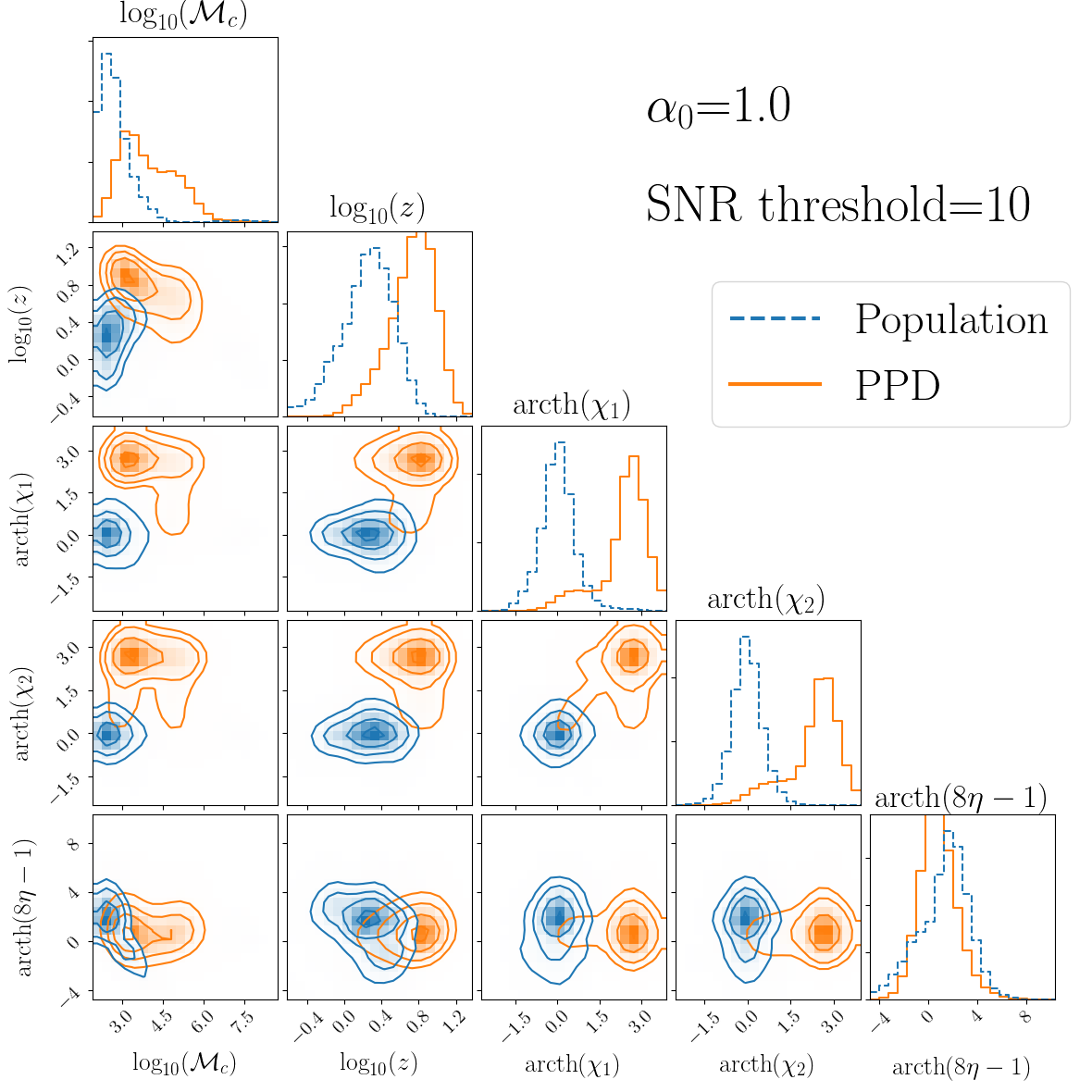}
%    }
\centering
%\subfigure[Intrinsic population.]{
    \centering \includegraphics[scale=0.22]{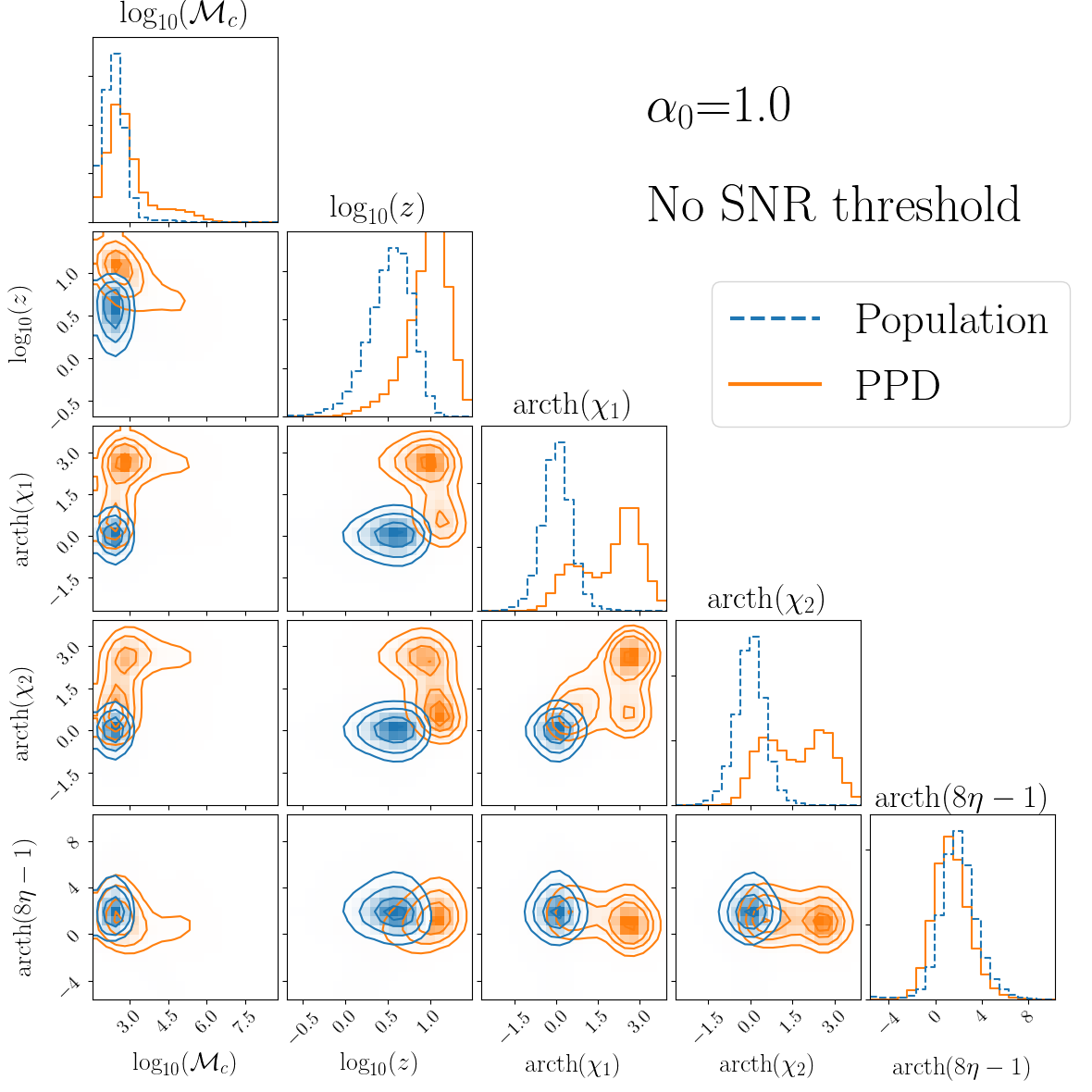}
%   }
   \centering
   \caption{Comparison between the population distribution for the SN-delays model and the PPD for an observation set containing 20 events from the same catalogue. Different rows refer to the HS variant ($\alpha_0=0$, top), a mixing fraction $\alpha_0=0.5$ between the HS and LS variants (middle), and the LS variant ($\alpha_0=1$, bottom). Panels on the left refer to the detectable population; panels on the right, to the intrinsic population. 
%We can reasonably well reproduce the chirp mass distribution of the detectable population, but we overestimate the fraction of light events in the intrinsic population.
}
\label{fig:ppd_2020}   
\end{figure*}

In order to have a more global view, we generate several observation sets with an increasing number of events, drawing the mixing fraction uniformly in $[0,1]$.
We estimate the shift on $\alpha$ as the difference between the maximum-posterior point $\alpha_{\rm max}$ and the injection value $\alpha_0$, and the error on the mixing fraction $\Delta \alpha$ as the 90\% confidence interval centered around the median value.
In Fig.~\ref{fig:bias_error} we plot these quantities for two selected values of the number of observed events.
The color scale indicates the value of the injected mixing fraction $\alpha_0$ for each observation set. As expected, both tend to decrease as we observe more events. Also, note that the points are equally distributed on both sides of the $\alpha_{\rm max}=\alpha_0$ line, indicating that there is little systematic bias in our analysis, as we would expect given that the models used to generate and analyze the data are consistent. We find that the error on $\alpha$ tends to be smaller for injected values close to 0 or 1, with even smaller errors in the former case, in agreement with our discussion on the shape of the posterior above. 

Next, we assess our ability to infer the population distribution from an observed dataset, using the PPD defined in Eq.~\eqref{ppd}. In order to make a quantitative comparison, we compute the Kullback-Leibler (KL) divergence~\cite{kullback1951} between them, defined as
\begin{equation}
 D_{\rm KL}=\sum_{\theta}p_1(\theta)\ln \left ( \frac{p_1(\theta)}{p_2(\theta)} \right ), \label{sbbhs:kl}
\end{equation}
with $p_1$ and $p_2$ the distributions we wish to compare.
In Fig.~\ref{fig:ppd_kl}, we plot the KL divergence between the PPD and the population distribution for datasets of 50 and 200 observed events, taking the population distribution as the reference distribution ($p_1$). Given the similarity between the distributions (as indicated by the smallness of the KL divergence), the results would not be significantly altered had we chosen the PPD as the reference distribution. The KL divergence tends to be smaller for larger datasets, meaning that our inference on the population distribution improves. As a trend, the largest values of the KL divergence correspond to $\alpha_0 \sim 0$. This is because the population distribution varies faster for small $\alpha$, so even small (statistical) deviations in the estimation of the mixing fraction lead to larger discrepancies between the PPD and the population distribution for $\alpha_0 \sim 0$. As an illustration, in Fig.~\ref{fig:ppds} we compare the PPD obtained from four simulated LISA datasets of 100 observed events generated with different values of $\alpha_0$ to the corresponding population distribution. Those realizations are chosen to span the range of values of KL divergences. As can be seen in the upper-left panel, even in the worst case (the largest value of the KL divergence among the cases shown in Fig.~\ref{fig:ppd_kl}) we can reconstruct the population distribution reasonably well. The other panels show the comparison between the PPD and the population distribution for datasets of 100 events yielding mid-range values of the KL divergence and for the dataset yielding the smallest one. Overall, this pipeline allows us to infer the population distribution accurately when the model used to generate the data is the same as the one used in the pipeline. We will now test the robustness of this pipeline by using different models in the two stages.

\subsection{Robustness}
 \label{sec:results_b}

We mix the HS and LS variants of the SN-delays model as described in Eq.~\eqref{mix}, and generate datasets of 20 observed events for $\alpha_0=0$, $\alpha_0=0.5$ and $\alpha_0=1$. We run our pipeline on these observation sets, still using our fiducial astrophysical model in the hierarchical Bayesian analysis and compare the PPD to the population distribution. The results are shown in Fig.~\ref{fig:ppd_2020}. In each case, we show both the intrinsic distribution and the detected one (where detection is defined by imposing an SNR threshold of 10). 
For $\alpha_0=0$ (top panels), we can reproduce reasonably well the chirp mass distribution of the detectable population, but we overestimate the fraction of small-$\mathcal{M}_c$ events in the intrinsic population. This is because the HS variant of the SN-delays model has a tail extending to lighter values than the HS variant of the fiducial model, as can be seen on Fig.~\ref{fig:comp_models}. Our pipeline compensates for this by adding events from the LS variant, and since only $\sim 25\%$ of LS events are detectable, the fraction of light events in the intrinsic population is overestimated.
Similarly, for $\alpha_0=0.5$ (middle panels) the PPD agrees reasonably well with the population distribution of the chirp mass for detectable events, but this time the fraction of light events in the intrinsic population is underestimated. This is due to the difference in the fraction of detectable events between the LS variant of our fiducial model and the SN-delays model (see Table \ref{tab:rates}). For a given number of detected light events, the latter predicts twice as many light events in the intrinsic population as our fiducial model. Finally, for $\alpha_0=1$ (bottom panels) even the chirp mass distribution of detectable events is badly estimated. This is due to a tail of heavy events predicted by the LS variant of the SN-delays model, which causes our pipeline to estimate $\alpha_0$ to be different from 1. In all three cases, due to the differences in the fiducial and SN-delays population, redshift and spin distributions are poorly reconstructed. 
 
These results show that this pipeline would lead to erroneous predictions if the population of MBHBs is too different from the one predicted by our astrophysical models. Note that in the LS SN-delays model we do not expect to observe 20 events even for a ten-year mission duration, but this does not change our previous conclusion.  
 
\subsection{Including measurement errors}
\label{sec:results_c}

We now wish to consider two sources of error: weak lensing and statistical errors due to detector noise. They are accounted for with the following procedure. For each event predicted by the model:
\begin{itemize}
    \item[(1)] we draw a new value of the luminosity distance from a Gaussian distribution centered at the original value with variance given by the lensing error of Eq.~\eqref{lensing}, keeping the detector-frame mass constant;
    
    \item[(2)] from that new event, we draw a shifted event from a multinormal Gaussian distribution with covariance given by the Fisher information matrix at that point;
    
    \item[(3)] if this new event has SNR above the threshold, we perform parameter estimation;
    
    \item[(4)] we broaden the posterior distribution of the luminosity distance (and therefore of the redshift and the source frame mass) with the lensing error of Eq.~\eqref{lensing}. 
    
\end{itemize}
For step (3), we use the Fisher information matrix instead of doing a full Bayesian analysis in order to speed up computations. Some events from the LS variant have very low SNR of order unity, and in those cases the Fisher information matrix is poorly conditioned. For this reason, events with such low SNRs might end up with large enough SNRs to be detected after applying the Fisher matrix shift of step (2). This is not physically realistic, since the detector noise is unlikely to make such events detectable, and therefore between steps (2) and (3) we discard all events that have SNR below 5 {\em before} the shift. 

\begin{figure}[hbtp]
\centering
 \includegraphics[scale=0.1]{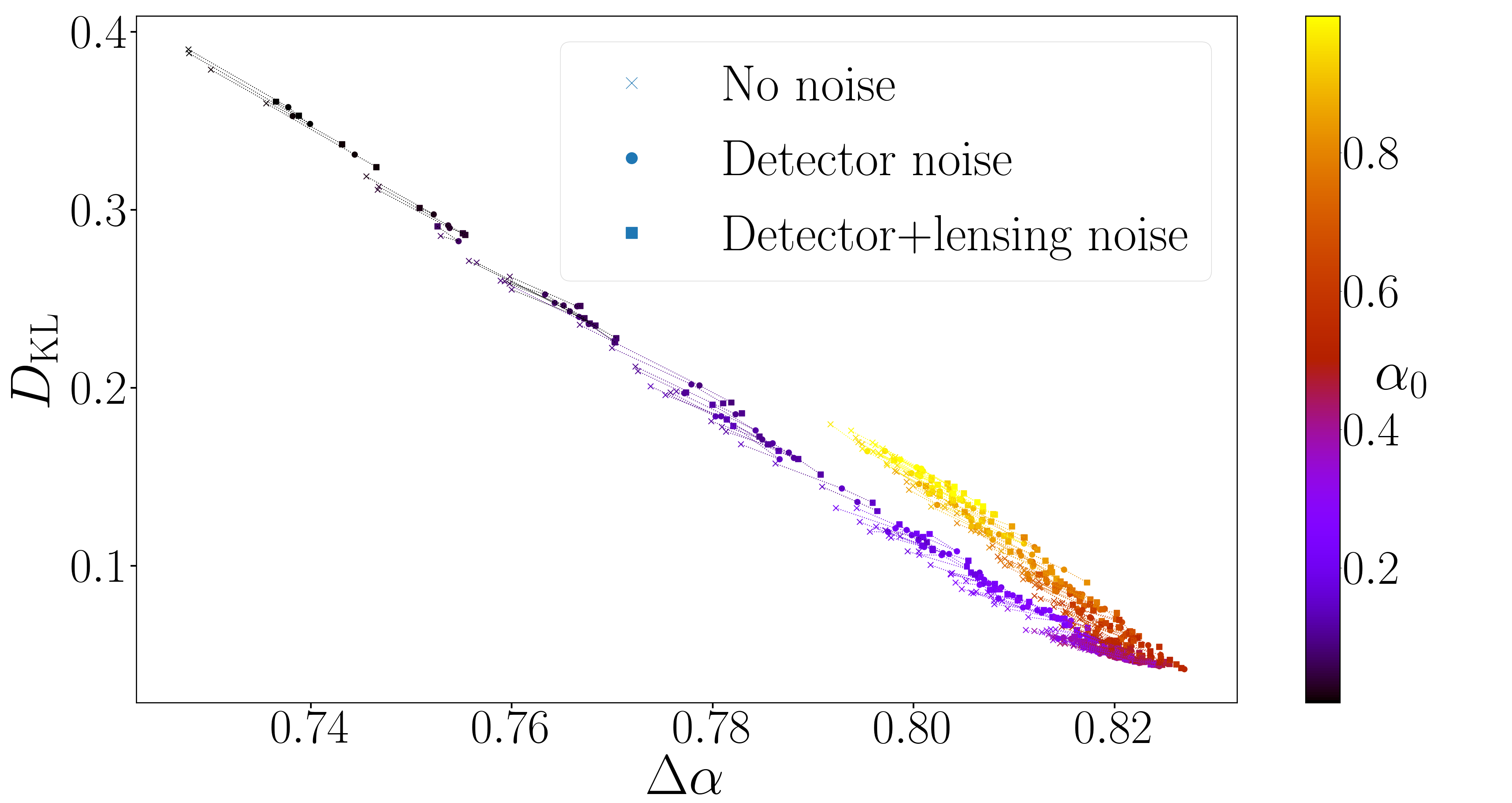}\\
 \caption{Error on $\alpha$ and KL divergence between the rescaled posterior distribution of $\alpha$ and the (flat) prior. As we include the different sources of error, $\Delta \alpha$ tends to increase and $D_{KL}$ tends to decrease, reflecting a degradation in the measurement of $\alpha$. Note that these are the errors and KL divergences for the rescaled posterior, i.e.~we artificially bring the number of detected events to 1, as detailed in the main text.}\label{fig:noises}
\end{figure}

In order to assess the impact of measurement errors, we generate datasets of 500 events (before applying the detectability criterion) for $\alpha_0$ drawn randomly in $[0,1]$, and consider three scenarios:
\begin{itemize}
    \item[(i)] there is no noise, i.e., none of the steps above are applied;
    \item[(ii)] there is only detector noise, i.e., only steps (2) and (3) are applied;
    \item[(iii)] there is both detector noise and lensing noise, i.e. all four steps are applied.
\end{itemize}
Note that steps (1) and (2) modify the number of detectable events, therefore we have to include these effects in the computation of the selection function. Moreover, increasing the number of observed events tends to narrow the posterior distribution, so in order to scale out this effect and allow for a fair comparison between the three different scenarios, we define a "rescaled" posterior distribution $\tilde{p}(\alpha|{\rm d})=p(\alpha|{\rm d})^{1/N_{\rm obs}}$. In Fig.~\ref{fig:noises} we plot on the $x$-axis the error on $\alpha$ (obtained from the rescaled posterior) and on the $y$-axis the KL divergence between the rescaled posterior distribution of $\alpha$ and the (flat) prior on $\alpha$, for different datasets and in the three scenarios. The color scale indicates the value of $\alpha_0$. The larger the KL divergence, the more information we gain from the dataset. As expected, including the different sources of error tends to decrease the KL divergence and increase $\Delta \alpha$. The dotted lines going from the top-left to bottom-right link simulations with the same underlying populations, and show (slight) degradation in the measurement of $\alpha$. Note that the KL divergence is larger and the error smaller for $\alpha_0\sim0$ and also for $\alpha_0\sim1$, in agreement with the discussion on the shape of the posterior in the previous subsection. Finally, we do not observe the appearance of systematic biases when including measurement errors.

\begin{figure}[hbtp]
\centering
 \includegraphics[scale=0.3]{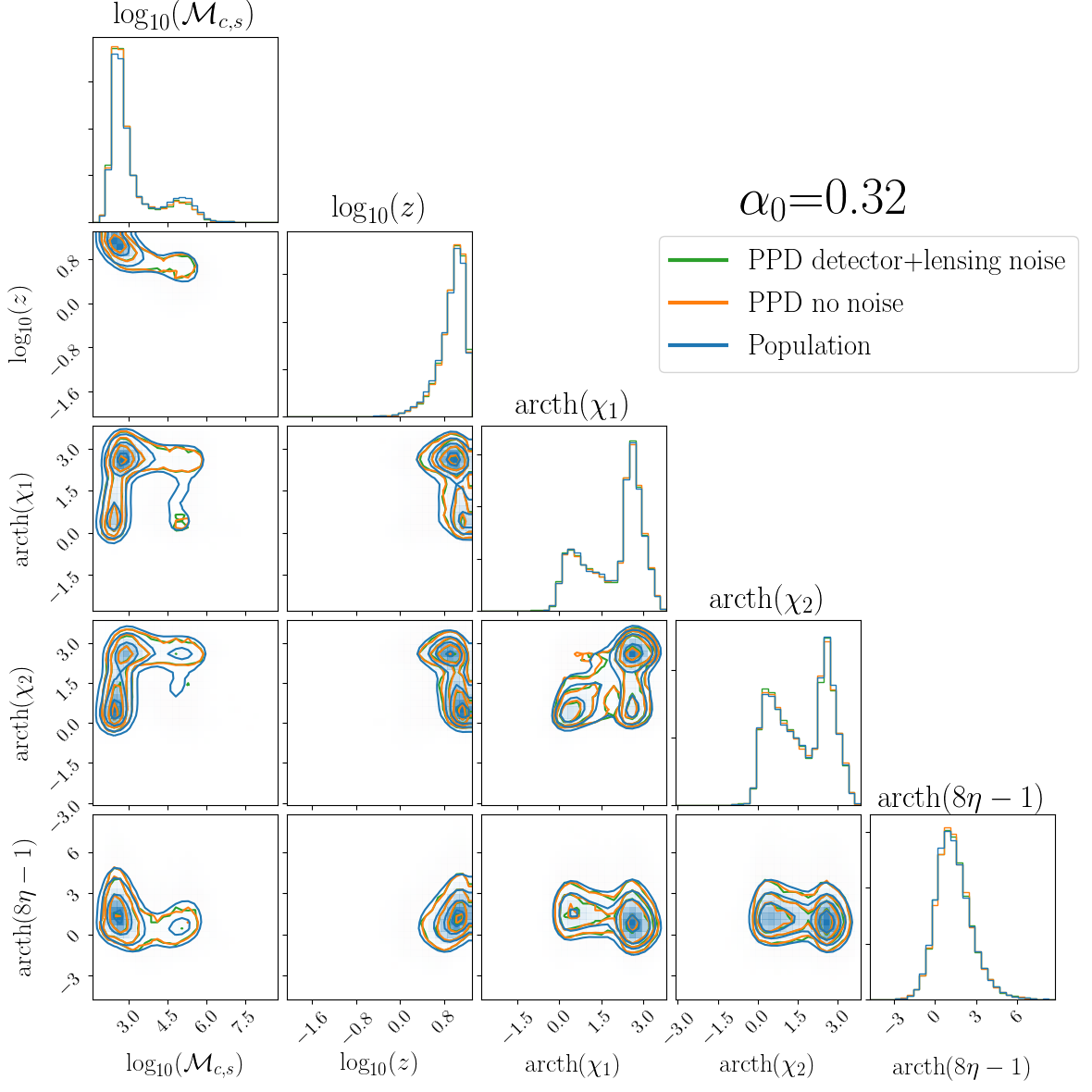}\\
 \caption{Population distribution and PPDs obtained in the no-noise and detector+lensing noise scenarios for a representative case. The dataset contains 500 events (before applying the detectability criterion). Including measurement errors barely affects our ability to infer the population distribution.}\label{fig:ppd_noises}
\end{figure}

Although the determination of $\alpha$ gets slightly worse when including the different sources of error, this barely affects our inference of the population distribution of MBHBs, as can be seen in Fig.~\ref{fig:ppd_noises}. There, we compare for a representative case the population distribution to the PPDs obtained in the no-noise and in the detector+lensing noise scenarios, which look very similar. 

Finally, we performed a last test: we generated datasets including both sources of noise in steps (1) and (2), but we did not include the effect of lensing in the hierarchical analysis, i.e. step (4). Moreover, we used the selection function obtained when accounting only for the detector noise. Our goal is to assess how our analysis would be biased if we did not properly model the effect of lensing. We observe a tendency to bias the measurement of $\alpha$ toward higher values, but no real impact on the PPD. This could be an artifact of our simplistic model, and should be verified through further work. 

\section{Conclusions}\label{ccls}

In this paper we discussed of the ability of LISA to distinguish between different astrophysical models for the formation and evolution of MBHs by inferring the population of MBHBs. We introduced a mixing fraction between astrophysical models to account for the possibility that the population of MBHBs in the Universe cannot be described by one single model. More specifically, we mixed between two variants of the same model: one that predicts that MBHs form from LSs and another from HSs. We built a pipeline based on the hierarchical Bayesian framework to measure the mixing fraction from LISA observations, and infer the population of MBHs. We have shown that this pipeline allows us to reconstruct accurately the population of MBHBs if it is similar to the one used in the pipeline, but not if the populations are too different. 
 
This problem could be mitigated by introducing more flexibility in the population model, at the cost of having greater uncertainty in the inferred population distribution. One approach would potentially be to include additional mixing fractions: one could in principle mix between as many models as desired. However, given the large uncertainty surrounding astrophysical models, we believe a better alternative is to use a theory-agnostic approach. We are currently working on a simplified astrophysical model for the formation and evolution of MBHs where the population of MBHBs depends on physically meaningful hyperparameters controlling the initial mass distribution, the delay between dark matter halo mergers and MBHB mergers, etcetera. We could then perform a hierarchical Bayesian analysis to infer these hyperparameters from LISA observations.
%using this same model to generate data, or more complex ones, such as the ones we used in this work. \eb{last sentence is unclear}
 
We have shown that measurement errors due to lensing and detector noise will not significantly impact our ability to infer the MBHB population. On the other hand, mismodelling the effect of weak lensing could lead to biases in our analysis. In our model, this bias has a negligible impact on our inference of the population of MBHBs, but this could be due to the simplicity of our model and will have to be further verified for different models. Finally, we commented on an important aspect: analyses based on results from numerical simulations, such as ours, require a large number of points in order to properly evaluate the probability density function of the theoretical model and the selection function, and thus avoid systematic biases. We estimate that at least a few tens of thousands of points are needed. 

Concerning our astrophysical model, we mixed the distributions {\it a posteriori}, i.e. with the results obtained by running numerical simulations with LSs and HSs independently. Therefore, our model cannot account for mergers between BHs formed from LSs and HSs and how they impact the population distribution. This could be included by mixing the seeding prescriptions {\it a priori}, when running the simulations. We could then use these results to assess the validity of our {\it a posteriori} approach. We leave this for future work.

\begin{acknowledgments}
%%%%%%%%%%%%%%%%%%%%%%%%%%%%%%%%%%%%%%%%%%%%%%%%%%%%%%%%%%%%%%%%%%%%%%%%%%%%%
%%%%%%%%%%%%%%%%%%%%%%%%%%%%%%%%%%%%%%%%%%%%%%%%%%%%%%%%%%%%%%%%%%%%%%%%%%%%%
A.T. is thankful to Johns Hopkins University for their hospitality during the early stages of this work. 
 S.B. and A.T. acknowledge support by CNES, in the framework of the LISA mission. 
This work has been supported by the European Union's Horizon 2020 research and innovation program under the Marie Sk\l{}odowska-Curie grant agreement No 690904.
E.~Barausse acknowledges financial support provided under the European Union's H2020 ERC Consolidator Grant ``GRavity from Astrophysical to Microscopic Scales'' grant agreement no. GRAMS-815673.
E.~Berti and K.W.K.~Wong are supported by NSF Grants No. PHY-1912550 and No. AST-2006538, NASA ATP Grants No. 17-ATP17-0225 and No. 19-ATP19-0051, NSF-XSEDE Grant No. PHY-090003, and NSF Grant No. PHY-20043. This research project was conducted using computational resources at the Maryland Advanced Research Computing Center (MARCC).
S.~Taylor is supported by NSF Grant No. AST-2007993 and PHY-2020265.
The authors would like to acknowledge networking support from the COST Action CA16104.
 \end{acknowledgments}

%\newpage
 
 \appendix
 
 \section{Kernel density estimation}\label{app:kde}

\begin{figure}[t]
\centering
 \includegraphics[scale=0.08]{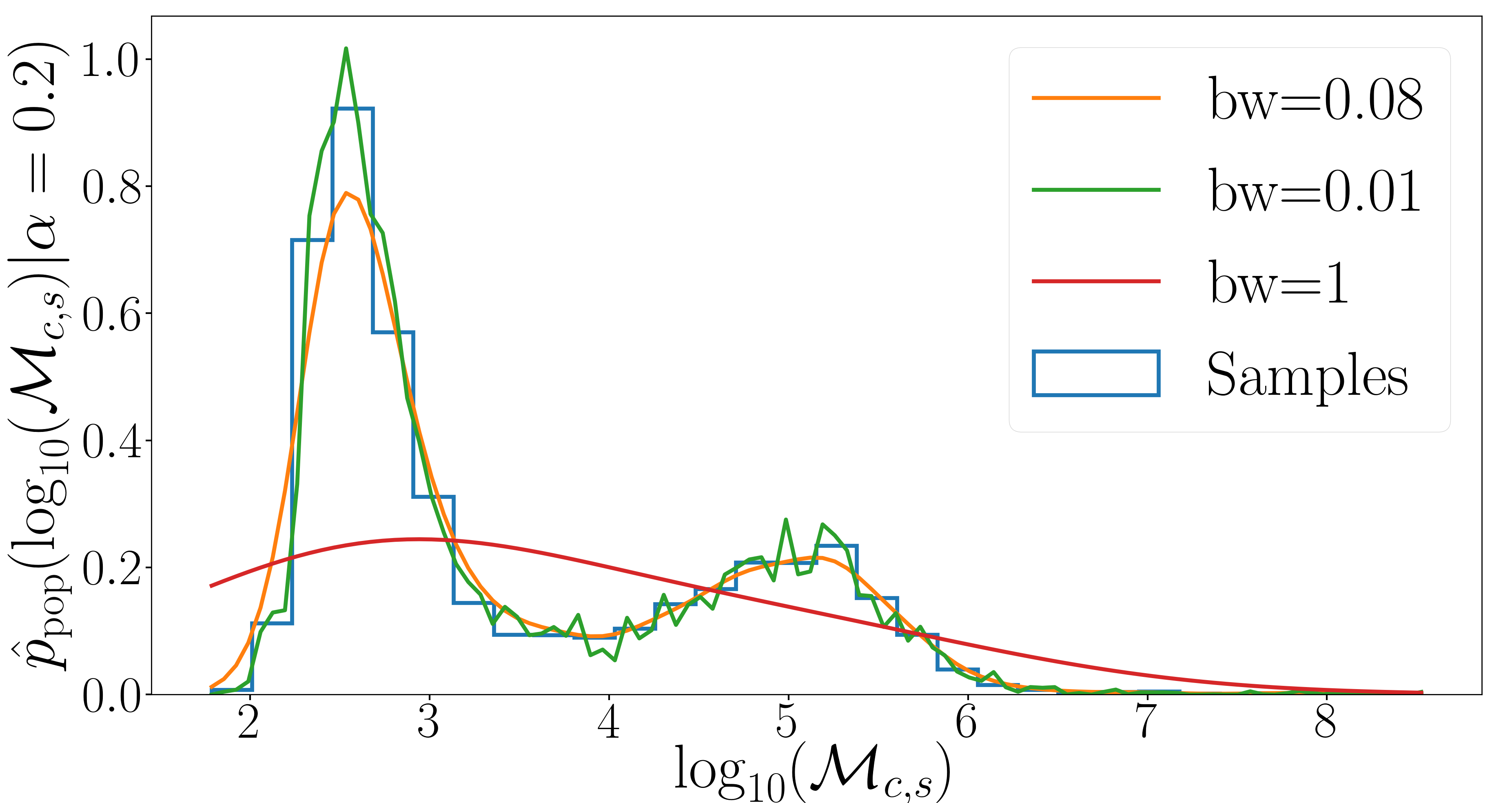}\\
 \centering
 \caption[Comparison between different KDE approximations to the population probability density function of $\log_{10}(\mathcal{M}_{c,s})$, using different values of the bandwidth.]{Comparison between different KDE approximations to the population probability density function of $\log_{10} \mathcal{M}_{c,s}$, using different values of the bandwidth. If the bandwidth is too small the KDE is not smooth, and if it is  too large we cannot resolve the features of the distribution. For the case shown here, a bandwidth of 0.08 is a good choice. This value was obtained by minimizing the integrated squared error, as described in the main text.}\label{comp_kdes}
\end{figure}

From a set of $n_s$ samples drawn from the distribution $p_{\rm pop}(\theta|\alpha)$, the KDE approximates its probability density function as 
\begin{equation}
 \hat{p}_{\rm pop}(\theta|\alpha)=\frac{1}{n_s} \sum_{i=1}^{n_s}K_{H}(\theta-\theta_i),
\end{equation}
where $K_H$ is the {\it kernel function}. We choose to work with Gaussian KDEs, where, denoting by $n_d$ the dimensionality of the parameter space,
\begin{equation}
 K_{H}(y)=\frac{1}{(2\pi)^{n_d/2}} [{\rm det}(H)]^{-1/2}e^{-\frac{1}{2}y^{T}H^{-1}y}.
\end{equation}
In the Gaussian KDE implementation of \texttt{scipy}~\cite{2020SciPy-NMeth}, $H$ is taken to be proportional to the identity matrix. The proportionality constant is called the {\it bandwidth} of the KDE, and is a very important parameter, since it defines the smoothing scale of the approximation to the target probability density function. In Fig.~\ref{comp_kdes} we show the approximations to the population probability density function of $\log_{10}(\mathcal{M}_{c,s})$ that we obtain using different values of the bandwidth (noted bw). 

For too large values of the bandwidth, we cannot resolve the features of the distribution, and for too small values, the resulting probability density function is not smooth. We deal with this issue by choosing the bandwidth that minimizes the integrated squared error $\int {\rm d} \theta (p_{\rm pop}(\theta|\alpha) -\hat{p}_{\rm pop}(\theta|\alpha))^2$. In practice, it is estimated by using a Monte Carlo averaging, and the quantity we seek to minimize is~\cite{10.1214/aos/1176348376}
\begin{equation}
 \int {\rm d} \theta \hat{p}_{\rm pop}(\theta|\alpha)^2 - \frac{2}{n_s}\sum_{i=1}^{n_s} \hat{p}_{\rm pop,-i}(\theta_i|\alpha), 
\end{equation}
where the sum runs over the $n_s$ samples drawn from $p_{\rm pop}(\theta|\alpha)$ used to approximate the integral, and $\hat{p}_{\rm pop,-i}(\theta|\alpha)$ is the KDE obtained using all $n_s$ samples but the $i^{th}$ one. The value of 0.08 used in Fig.~\ref{comp_kdes} was obtained with this method. We also apply it to compute the bandwidth of the KDE for the LS and HS population distributions. 
 
\section{Systematic biases due to misevaluation of the selection function}\label{app:selection} 

\begin{figure}[hbtp!]
 \centerline{\includegraphics[scale=0.08]{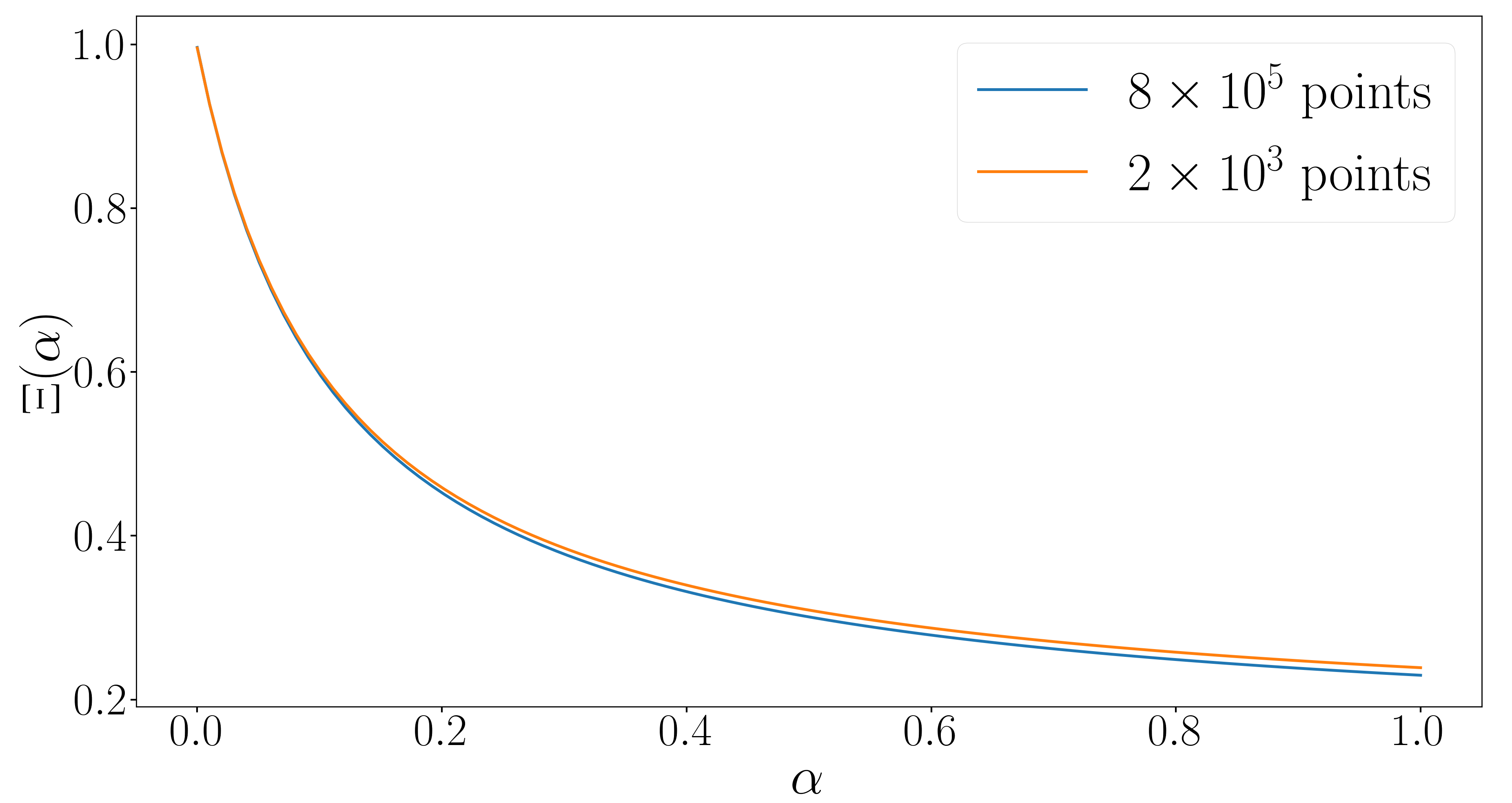}}
 \caption[Comparison between the selection functions obtained using different numbers of points.]{Comparison between the selection functions obtained using different numbers of points.}\label{app:fig:comp_selection}
 \end{figure}

\begin{figure}[hbtp!]
 \centering
\subfigure[We use $8\times 10^5$ points to evaluate the selection function of the LS and HS variants.]{
    \centering \includegraphics[scale=0.1]{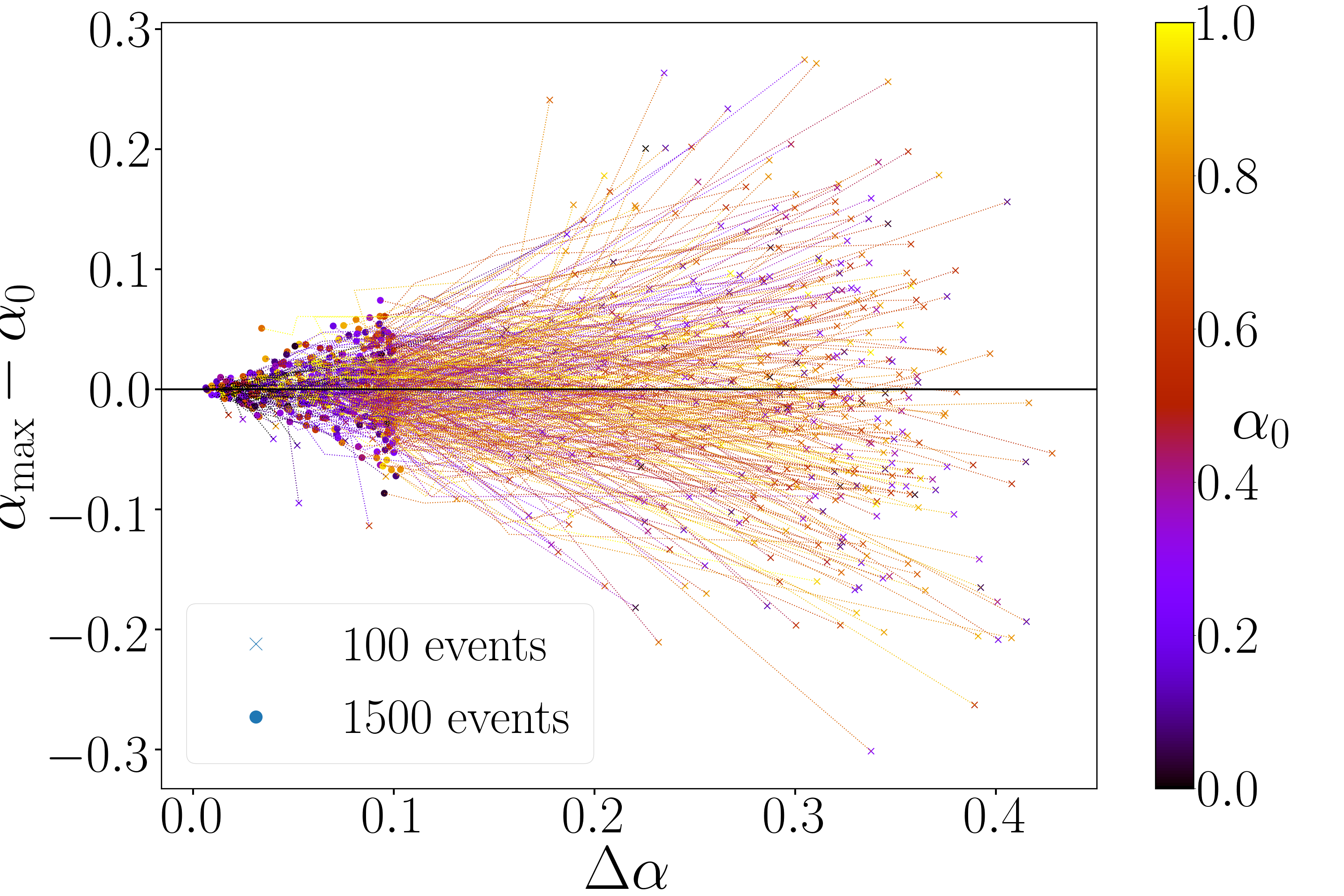}
    }
\centering
\subfigure[We use $2\times 10^3$ points to evaluate the selection function of the LS and HS variants.]{
    \centering \includegraphics[scale=0.1]{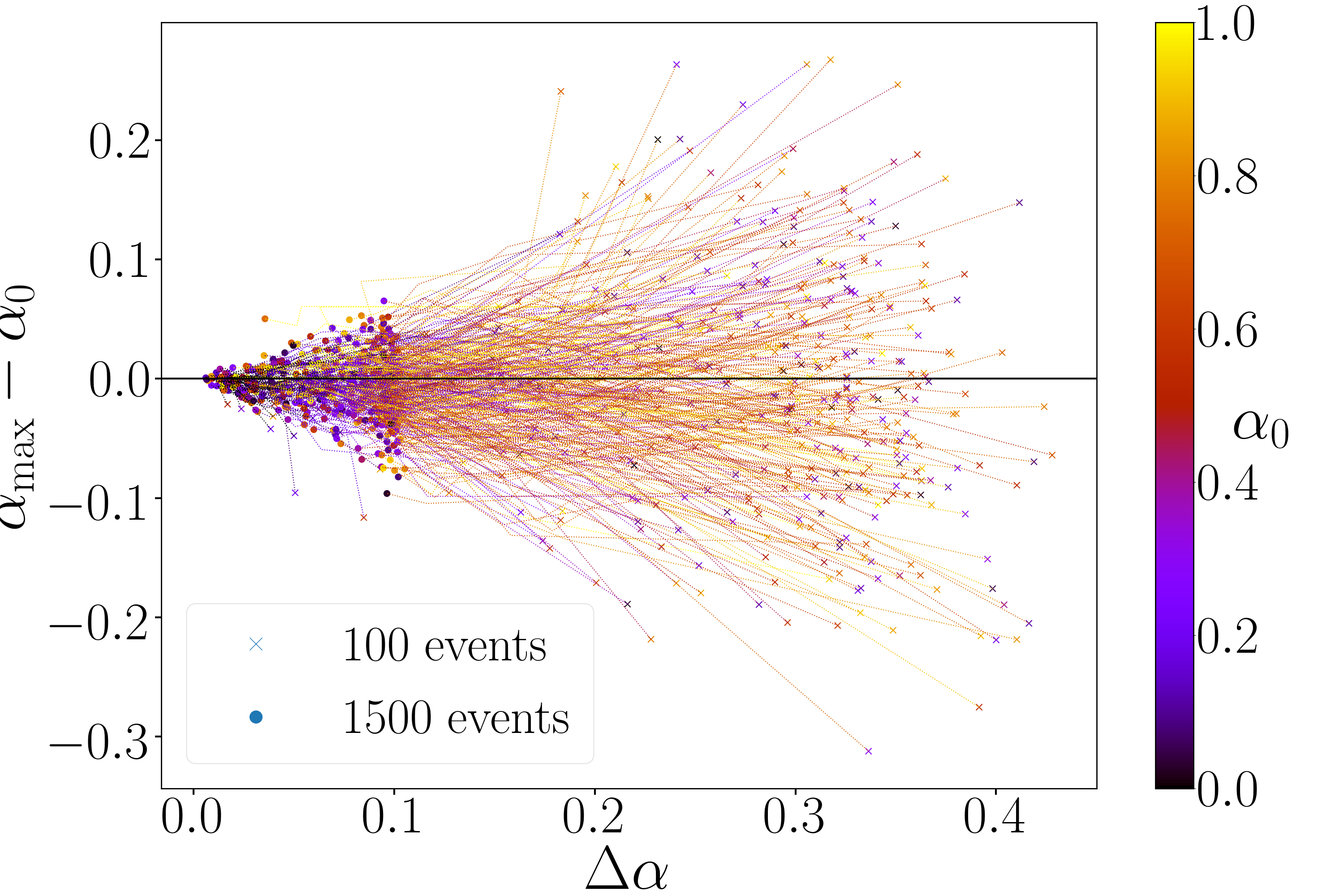}
   }
   \centering
   \caption[]{Evolution of the bias and the error on $\alpha$ using the selection function in blue in Fig.~\ref{app:fig:comp_selection} (top) and the one in orange (bottom). We can clearly observe a systematic bias in the latter case due to misevaluation of the selection function.}\label{app:fig:bias_error_selections}   
 \end{figure}

The selection function used to obtain the results of this paper was computed with Eq.~\eqref{selection}. We generated $8 \times 10^5$ events for the LS and HS variants from the KDE and computed the terms $\Xi({\rm LS})$ and $\Xi({\rm HS})$ individually. In Fig.~\ref{app:fig:comp_selection} we compare this selection function with one obtained using only $2\times 10^3$ points to compute each term. There is a clear discrepancy between the two functions, which reflects on the population inference as can be seen in Fig.~\ref{app:fig:bias_error_selections}. There we compare the shift versus error on $\alpha$ plots obtained using each of these selection functions. Clearly, using too few points to compute the selection function leads to systematic biases, as can be seen by the fact that many more points are below the $\alpha_{\rm max}=\alpha_0$ line than above. We do not expect to observe thousands of MBHBs with LISA, but we have chosen this large number of events to emphasize this effect. Even for fewer events we could be biased due to misevaluation of the selection function, and a large number of points from numerical simulations will be needed to mitigate this effect (see also~\cite{Farr_2019}). Moreover, third generation ground-based detectors are expected to detect thousands of events, and will face this same problem. In our study, this systematic bias becomes negligible when using $\mathcal{O}(10^5)$ points for each model.

\section{Comparison between KDE and the population obtained from simulations}
\label{app:modelcomp}

\begin{figure}[t]
% \centering
%\subfigure[LS variant.]{
    \centering \includegraphics[scale=0.27]{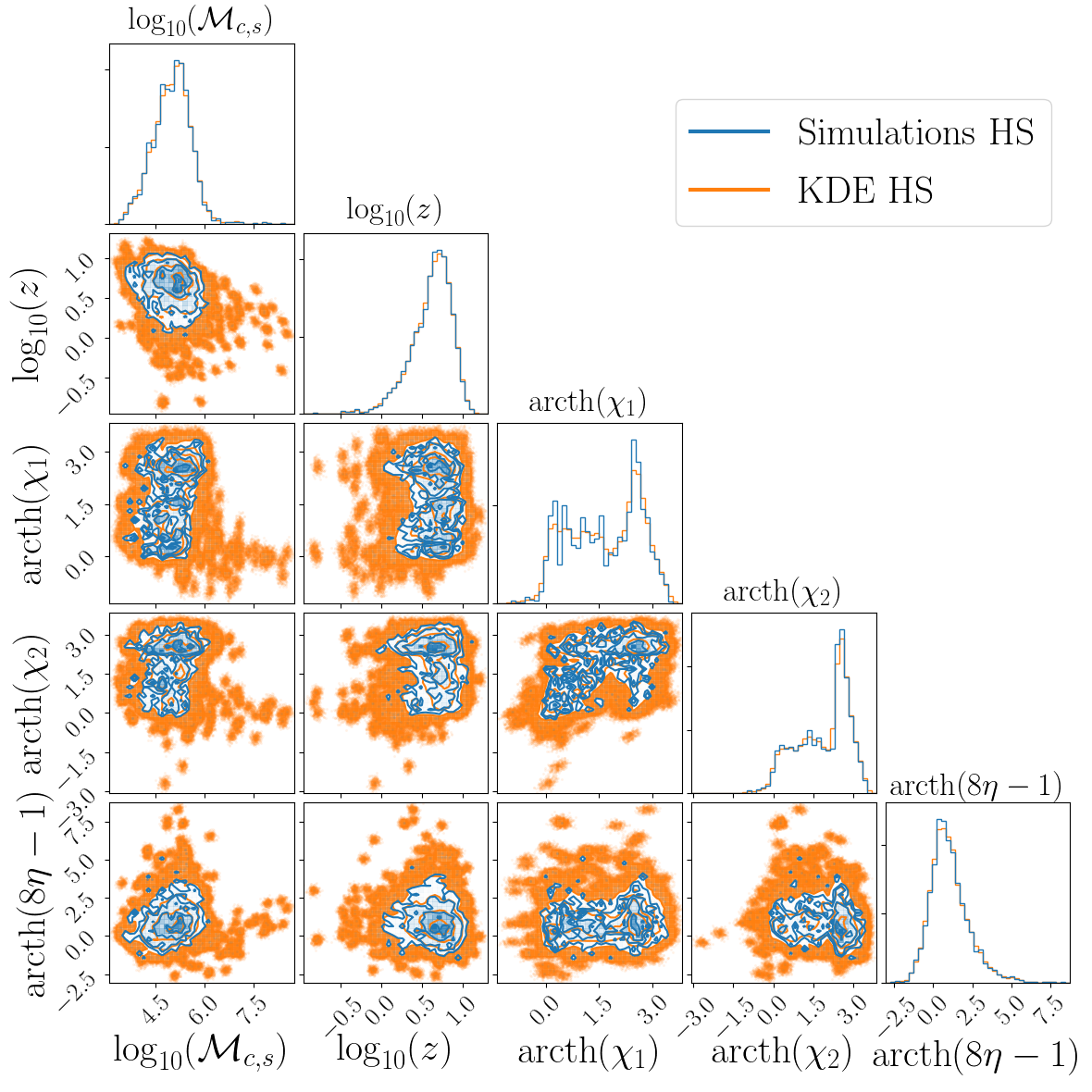}
%    }
%\centering
%\subfigure[HS variant.]{
    \centering \includegraphics[scale=0.27]{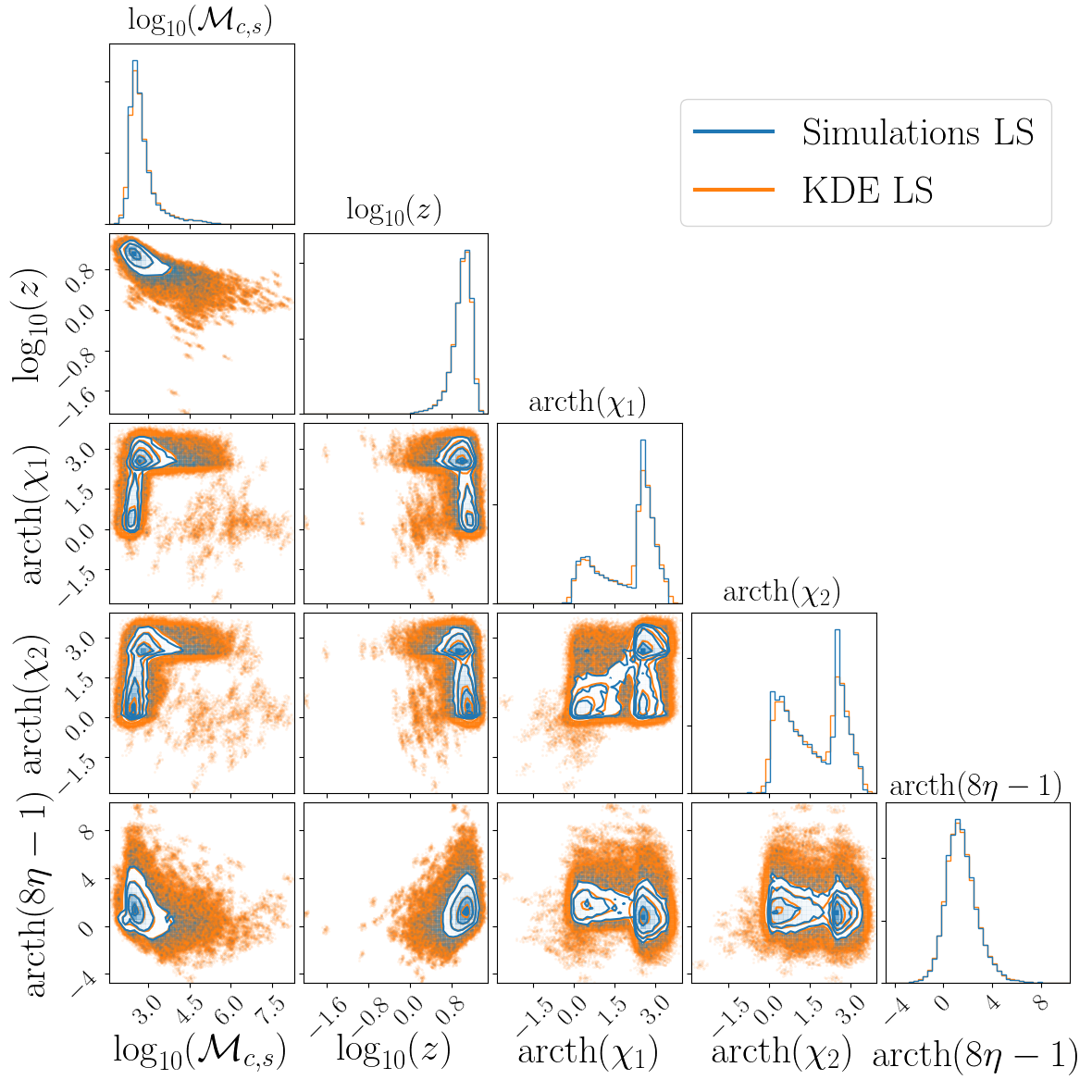}
%   }
%   \centering
   \caption{Comparison between the population distributions obtained from numerical simulations and the KDE we build from it. We purposefully did not smooth the corner plot in order to reflect the real level of agreement between the two distributions. The top and bottom panels refers to the LS and HS variants, respectively. The ``bumpy'' histograms for the HS variant (in particular for the spin) highlight that we do not have enough points to build an accurate enough KDE for our purposes. However the two distributions are overall in good agreement, and therefore we expect that the approximation of using the KDE as our ``true'' fiducial astrophysical model should not sensibly affect our results.}\label{fig:comp_cat_kde}   
\end{figure}

In Fig.~\ref{fig:comp_cat_kde} we compare the population distribution predicted from numerical simulations to the one obtained from building a KDE on it. 

  \FloatBarrier
\bibliography{Ref}
\end{document}